\documentclass[12pt,cls,onecolumn]{IEEEtran}
\usepackage{graphicx,amsmath,amssymb,epsfig, amsfonts, cite, latexsym, cuted, multicol, multirow, subfigure, stfloats, array, tabularx}
\usepackage{subeqnarray,tabularx}
\usepackage{setspace}
\usepackage{anysize}
\usepackage{subfigure,cite,graphicx,amsmath,amssymb,mathrsfs,epsf}
\usepackage[usenames]{color}

\begin{document}

\title{Capacity of Three-Dimensional Erasure Networks}
\author{\large Cheol Jeong, {\em Member}, {\em IEEE}, and Won-Yong
Shin, {\em Senior Member}, {\em IEEE}
\\
\thanks{This research was
supported by the Basic Science Research Program through the
National Research Foundation of Korea (NRF) funded by the Ministry
of Science, ICT \& Future Planning (MSIP) (2015R1A2A1A15054248).}
\thanks{C. Jeong is with the DMC R\&D Center,
Samsung Electronics, Suwon, Republic of Korea. (E-mail:
cheol.jeong@ieee.org).}
\thanks{W.-Y. Shin is with the Department of Computer Science and
Engineering, Dankook University, Yongin 448-701, Republic of Korea
(E-mail: wyshin@dankook.ac.kr).}
} \maketitle


\markboth{IEEE Transactions on Communications} {Jeong and Shin:
Capacity of Three-Dimensional Erasure Networks}


\newtheorem{definition}{Definition}
\newtheorem{theorem}{Theorem}
\newtheorem{lemma}{Lemma}
\newtheorem{example}{Example}
\newtheorem{corollary}{Corollary}
\newtheorem{proposition}{Proposition}
\newtheorem{conjecture}{Conjecture}
\newtheorem{remark}{Remark}

\def \diag{\operatornamewithlimits{diag}}
\def \min{\operatornamewithlimits{min}}
\def \max{\operatornamewithlimits{max}}
\def \log{\operatorname{log}}
\def \max{\operatorname{max}}
\def \rank{\operatorname{rank}}
\def \out{\operatorname{out}}
\def \exp{\operatorname{exp}}
\def \arg{\operatorname{arg}}
\def \E{\operatorname{E}}
\def \tr{\operatorname{tr}}
\def \SNR{\operatorname{SNR}}
\def \dB{\operatorname{dB}}
\def \ln{\operatorname{ln}}

\def \bmat{ \begin{bmatrix} }
\def \emat{ \end{bmatrix} }

\def \be {\begin{eqnarray}}
\def \ee {\end{eqnarray}}
\def \ben {\begin{eqnarray*}}
\def \een {\end{eqnarray*}}

\begin{abstract}
In this paper, we introduce a large-scale three-dimensional (3D)
erasure network, where $n$ wireless nodes are randomly distributed
in a cuboid of $n^{\lambda}\times n^{\mu}\times n^{\nu}$ with
$\lambda+\mu+\nu=1$ for $\lambda,\mu,\nu>0$, and completely
characterize its capacity scaling laws. Two fundamental path-loss
attenuation models (i.e., exponential and polynomial power-law
models) are used to suitably model an erasure probability for
packet transmission. Then, under the two erasure models, we
introduce a routing protocol using percolation highway in 3D
space, and then analyze its achievable throughput scaling laws. It
is shown that, under the two erasure models, the aggregate
throughput scaling $n^{\min\{1-\lambda,1-\mu,1-\nu\}}$ can be
achieved in the 3D erasure network. This implies that the
aggregate throughput scaling $n^{2/3}$ can be achieved in 3D {\it
cubic} erasure networks while $\sqrt{n}$ can be achieved in
two-dimensional (2D) {\it square} erasure networks. The gain comes
from the fact that, compared to 2D space, more {\it geographic
diversity} can be exploited via 3D space, which means that
generating more simultaneous percolation highways is possible. In
addition, cut-set upper bounds on the capacity scaling are derived
to verify that the achievable scheme based on the 3D percolation
highway is order-optimal within a polylogarithmic factor under
certain practical operating regimes on the decay parameters.
\end{abstract}


\newpage

\section{Introduction}

he Internet evolves into the next phase: the network consisting of
smart devices equipped with sensors and radio frequency
transceivers, connected to the Internet for sharing information
with each other, which is known as the Internet of Things (IoT).
Such smart devices are used for smart home, smart building, smart
metering, etc. For example, home appliances are connected to
wireless networks so that a user controls its home appliances
remotely. In a smart building, heating, ventilation, air
conditioning, and lighting are automatically controlled. Utilizing
a smart metering where utility meters such as electricity, water,
and gas are reported to the data center, not only customers can
save energy but also energy suppliers can help better serve their
customers. Machine-to-machine (M2M) or machine-type
communications~\cite{LoLawJacobsson:WC13} has recently received a
lot of attention as an enabling technology of IoT. As a massive
number of devices participate in M2M communications, a study on
the capacity scaling law has been taken into account as one of the
most challenging issues in understanding a fundamental limit on
the network throughput and its asymptotic trend with respect to
the number of devices. One can obtain remarkable insights into the
practical design of a protocol by characterizing such a capacity
scaling behavior.

\subsection{Previous Work}
In~\cite{GuptaKumar:00}, the throughput scaling for Gaussian
channels was originally introduced and characterized in a large
wireless ad hoc network where nodes are distributed in
two-dimensional (2D) space. It was shown that the aggregate
throughput scales as $\Theta(\sqrt{n/\log n})$ in a network having
$n$ nodes randomly distributed in a unit area.\footnote{We use the
following notation: i) $f(x)=O(g(x))$ means that there exist
positive constants $B$ and $b$ such that $f(x)\leq Bg(x)$ for all
$x>b$, ii) $f(x)=o(g(x))$ means that $\lim_{x\rightarrow
\infty}\frac{f(x)}{g(x)}=0$, iii) $f(x)=\Omega(g(x))$ if
$g(x)=O(f(x))$, iv) $f(x)=w(g(x))$ if $g(x)=o(f(x))$, and v)
$f(x)=\Theta(g(x))$ if $f(x)=O(g(x))$ and $g(x)=O(f(x))$
\cite{D.Knuth:76}.} This throughput scaling is achieved using the
nearest-neighbor multihop (MH) routing scheme (also known as the
Gupta--Kumar routing scheme). MH schemes were further studied and
analyzed
in~\cite{FranceschettiDouseTseThiran:07,GuptaKumar:03,ShinChungLee:TIT13,NebatCruzBhardwaj:09,ElGamalMammenPrabhakarShah:06,NeelyModiano:05}.
In~\cite{FranceschettiDouseTseThiran:07}, the aggregate throughput
scaling was improved to $\Theta(\sqrt{n})$ using percolation
theory, which newly models the connectivity of wireless networks.
It was shown that a hierarchical cooperation (HC)
strategy~\cite{OzgurLevequeTse:07,GhaderiXieShen:TIT09,NiesenGuptaShah:09}
achieves an almost linear throughput scaling, i.e.,
$\Theta(n^{1-\epsilon})$ for an arbitrarily small $\epsilon>0$, in
the Gaussian network model. In addition, in Gaussian networks of
unit area, there has been a lot of research to improve the
aggregate throughput up to a linear scaling by using novel
techniques such as networks with node
mobility~\cite{GrossglauserTse:02}, interference
alignment~\cite{CadambeJafar:08}, directional
antennas~\cite{ZhangXuWangGuizani:TC10,LiZhangFang:TMC11,YoonShinJeon:ISIT14},
and infrastructure
support~\cite{O.Dousse:INFOCOM02,KulkarniViswanath:03,KozatTassiulas:03,LiuLiuTowsley:03,ZemlianovVeciana:05,LiuThiranTowsley:07,ShinJeonDevroyeVuChungLeeTarokh:08}.

The aforementioned studies focused only on 2D wireless networks.
In cities such as Manhattan, where there exist a number of
skylines, however, all kinds of nodes such as sensors, meters,
appliances, and traffic lights can be located in three-dimensional
(3D) space. Hence, in such a scenario, assuming a 3D network is
rather suitable. A large-scale, low-cost wireless sensor network
testbed was deployed in Singapore and the inter-floor connectivity
was tested~\cite{DoddavenkatappaChanAnanda:ICST11}. The
Zigbee-based IoT was investigated in 3D
terrains~\cite{LinLeuLiWu:Elsevier2013}. It is thus envisioned
that 3D wireless networks will receive much more attention in M2M
communications. In 3D Gaussian networks, there have been a few
studies in the
literature~\cite{GuptaKumar:01,LiPanFang:TN12,FranceschettiMiglioreMinero:09,HuWangYangZhangXuGao:TC10}
regarding the throughput scaling analysis.
In~\cite{GuptaKumar:01}, the capacity was analyzed under both
Protocol and Physical Models when $n$ nodes are distributed in a
sphere. A more general physical model was employed
in~\cite{LiPanFang:TN12} to better characterize the throughput
from an information-theoretic point of view.
In~\cite{FranceschettiMiglioreMinero:09}, it was shown that
per-node throughput scales as $O((\log n)^3/n^{1/3})$ using
Maxwell's physics of wave propagation in a 3D network where nodes
are located inside a sphere. Similarly as in the 2D network
scenario~\cite{FranceschettiDouseTseThiran:07}, per-node
throughput scaling of $\Omega(1/n^{1/3})$ was shown to be achieved
using percolation theory in a cubic
network~\cite{HuWangYangZhangXuGao:TC10}.

Besides the Gaussian channel setup, another fundamental class of
channel models is an erasure network, originally introduced
in~\cite{DanaGowaikarHassibi:06}, where signals are either
successfully delivered or completely lost. This erasure channel
model is well suited for packetized systems where all information
in a packet may be lost due to the errors. The erasure channel
plays an important role in information theory and coding
theory~\cite{LeeUrbankeBlahut:08,JaberAndrews:11}. This is
because, by modelling each communication channel in wireless
networks as a memoryless erasure channel, one can easily tackle a
long-standing open problem, corresponding to the capacity region
for general Gaussian multiterminal networks
(see~\cite{TulinoVerduCaireShamai:07,VerduWeissman:08} and
references therein). In large-scale wireless networks, the erasure
probability will increase as the physical distance between one
transmitter and its intended receiver increases since the link
quality between the two nodes is degraded with distance. To
incorporate this phenomenon into the erasure channel model, one
can use either an exponential or polynomial decay model in
computing the path-loss attenuation. In 2D erasure networks, the
capacity scaling law was studied under the exponential decay
model~\cite{SmithGuptaViswanath:07ISIT,ShinKim:11}, while the
results
in~\cite{SmithVishwanath:06,SmithGuptaViswanath:07,JeongShin:CL13}
assumed the polynomial power-law model in analyzing the capacity
scaling law.

In~\cite{DanaGowaikarHassibi:06}, it was shown that network coding
at intermediate nodes is needed to achieve the capacity region in
a wireless erasure network, where each node transmits linear
combinations of the received non-erased symbols. However, network
coding does not further improve the throughput performance in
large-scale networks as long as scaling laws are
concerned~\cite{SmithGuptaViswanath:07,B.Smith:08}. In other
words, MH routing, which is a simple packet-forwarding scheme, is
order-optimal in large-scale erasure networks.
In~\cite{B.Smith:08}, the benefits of feedback were also
demonstrated, thereby allowing an extremely simple coding scheme.
In contrast to the network coding
in~\cite{DanaGowaikarHassibi:06}, a simple acknowledgement-based
feedback from the destination enables us to eliminate the
requirement for sending any side information including the erased
locations in a packet~\cite{B.Smith:08}. Adding such a feedback,
however, does not fundamentally change the throughput scaling law,
compared to the case where no feedback is allowed.

\subsection{Contributions}
In this paper, we introduce a general erasure network in 3D space,
where $n$ wireless ad hoc nodes are randomly distributed in a
cuboid of $n^{\lambda}\times n^{\mu}\times n^{\nu}$ with
$\lambda+\mu+\nu=1$ having unit node density for $\lambda, \mu,
\nu>0$. We also completely characterize its capacity scaling laws.
In the {\em Gaussian} channel model, both upper and lower bounds
on the capacity were shown in a 3D cubic
network~\cite{LiPanFang:TN12}, but it still remains open how to
further bridge the gap between the two bounds and then how to
characterize the capacity scaling. To the best of our knowledge,
the information-theoretic capacity scaling has never been analyzed
before for 3D {\em erasure} networks.

We start by introducing two fundamental path-loss attenuation
models (i.e., exponential and polynomial power-laws) to suitably
model an erasure probability for packet transmission. As
in~\cite{SmithGuptaViswanath:07ISIT,SmithGuptaViswanath:07},
inspired by the Physical model~\cite{GuptaKumar:00} under wireless
Gaussian networks, we use the finite-field additive interference
model in erasure networks. A {\em percolation highway} in 3D space
is then introduced to analyze our achievable throughput scaling
laws under the two practical erasure models. Specifically, a
highway system consisting of percolation highways in horizontal
and vertical directions over 2D networks is extended to the 3D
network configuration where there exist percolation highways in
three Cartesian directions. It is shown that, under both erasure
models, the aggregate throughput scaling
$n^{\min\{1-\lambda,1-\mu,1-\nu\}}$ can be achieved. This result
reveals that, in a cubic network (i.e.,
$\lambda=\mu=\nu=\frac{1}{3}$), the aggregate throughput scaling
$\Omega(n^{2/3})$ can be achieved under both erasure models. We
explicitly show how to operate our percolation-based highway
routing based on a {\em time-division multiple access (TDMA)}
scheme, which is not straightforward since finding the appropriate
number of required time slots is essential for obtaining the above
scaling result. We remark that this scaling is greater than the
aggregate throughput scaling $\Omega(\sqrt{n})$ achievable in 2D
square erasure networks. This is because, compared to 2D space,
more {\it geographic diversity} can be exploited via 3D
geolocation, which indicates that constructing more simultaneous
end-to-end percolation highways is possible. In addition, to
verify the optimality of the achievable scheme, we derive upper
bounds on the capacity scaling for each path-loss attenuation
model by using the max-flow min-cut theorem. It is shown that our
upper bound matches the achievable throughput scaling within a
polylogarithmic factor for all operating regimes under the
exponential decay model and for $\alpha>3$ under the polynomial
decay model, where $\alpha$ is the decay parameter for the
polynomial decay model. The capacity scaling law in a 3D erasure
network of unit volume (i.e., a dense network model) is also
characterized.

Our main contributions are threefold as follows:
\begin{itemize}
  \item We introduce a generalized 3D erasure network whose size is $n^{\lambda}\times n^{\mu}\times
n^{\nu}$ with $\lambda+\mu+\nu=1$ and propose a constructive
achievable scheme, i.e., a percolation-based 3D highway routing.
  \item We explicitly show our TDMA scheme by deriving the minimum number of required time slots to achieve the order optimality.
  \item We derive cut-set upper bounds on the capacity scaling, where both upper and lower bounds are of the same order within a polylogarithmic factor under a certain condition.
\end{itemize}

\subsection{Organization}
The rest of this paper is organized as follows. In
Section~\ref{SEC:SystemChannelModel}, the system and channel
models are described. The routing protocol based on percolation
highways is presented in Section~\ref{SEC:RoutingProtocol}.
Achievable throughput scaling laws are derived for both
exponential and polynomial decay models in
Section~\ref{SEC:Achievability}. Cut-set upper bounds on the
capacity scaling are then derived in Section~\ref{SEC:UpperBound}.
Extension to the dense network scenario is discussed in
Section~\ref{SEC:Dense}. Finally, we summarize the paper with some
concluding remark in Section~\ref{SEC:Conclusion}.

\section{System and Channel Models}\label{SEC:SystemChannelModel}

Consider a 3D wireless network where $n$ wireless nodes are
uniformly and independently distributed in a cuboid of
$n^{\lambda}\times n^{\mu}\times n^{\nu}$ with $\lambda+\mu+\nu=1$
having unit node density for $\lambda,\mu,\nu>0$ (i.e., an
extended
network~\cite{FranceschettiDouseTseThiran:07,OzgurLevequeTse:07}).
We randomly pick source--destination pairings so that each node is
the destination of exactly one source.

The channel between any two nodes is modelled as a memoryless
erasure channel with erasure events over all channels being
independent. We consider two models of erasure probability
$\epsilon_{ki}$ between nodes $i$ and $k$: the exponential decay
model~\cite{SmithGuptaViswanath:07ISIT,ShinKim:11} and the
polynomial decay
model~\cite{SmithGuptaViswanath:07,JeongShin:CL13}. In the
exponential decay model such that the probability of successful
transmission decays exponentially with a distance between two
nodes, it follows that
\begin{align}\label{Eq:ErasureProb-Exponential}
\epsilon_{ki} = 1-\gamma^{d_{ki}},
\end{align}
where $0<\gamma<1$ is the decay parameter for the exponential
decay model and $d_{ki}$ is the distance between nodes $i$ and
$k$. As another important model, the erasure probability can be
modeled as a polynomial power-law model in which the probability
of successful transmission decays polynomially with a distance
between two nodes. In this case, the erasure probability is given
by
\begin{align}\label{Eq:ErasureProb-Polynomial}
\epsilon_{ki} = 1-\frac{1}{d_{ki}^{\alpha}},
\end{align}
where $\alpha>0$ is the decay parameter for the polynomial decay
model.

Due to the broadcast nature of the wireless medium, under the
finite-field additive interference
model~\cite{SmithGuptaViswanath:07ISIT,SmithGuptaViswanath:07},\footnote{While
it has not been clearly studied in the literature what
interference means for erasure channels, the finite-field additive
interference model that incorporates the broadcast property into
the erasure channel was introduced
in~\cite{SmithGuptaViswanath:07ISIT,SmithGuptaViswanath:07}.} the
received symbol at node $k$ is given by
\begin{align}
y_k = \sum_{i\in \mathcal{I}}\eta_{ki}x_i, \nonumber
\end{align}
where $\mathcal{I}$ is the set of simultaneously transmitting
nodes, $\eta_{ki}$ is a Bernoulli random variable that takes the
value $0$ with probability $\epsilon_{ki}$ or $1$ with probability
$1-\epsilon_{ki}$, and $x_i$ is a single symbol chosen at node $i$
from the finite-field alphabet $\mathcal{X}$. The output $y_k$ is
the sum of all unerased symbols. It is assumed that each erased
packet does not contribute to any interference since the packet
can be successfully decoded due to the long packet length and
robust channel coding if the received signal power is sufficiently
high.

If a packet is erased at a certain receiver, then it will be
retransmitted from the receiver to the desired transmitter to
ensure the successful packet delivery based on the
acknowledgements of an automatic repeat request (ARQ) protocol.
For analytical convenience, we do not incorporate such a feedback
mechanism into our setting since it does not fundamentally affect
the achievability results as long as scaling laws are concerned.

\section{Routing Protocol}\label{SEC:RoutingProtocol}

It was shown that the nearest-neighbor MH routing is order-optimal
under the polynomial decay model in 2D erasure
networks~\cite{SmithGuptaViswanath:07}. On the other hand, the
nearest-neighbor MH routing is not sufficient in achieving the
optimal capacity under the exponential decay model since the
probability of successful transmission decreases {\em
exponentially} in distance $d_{ki}$ between two nodes while
per-hop distance (i.e., the distance between nodes in adjacent
routing cells) increases logarithmically, thus resulting in a huge
throughput degradation. Alternatively,
in~\cite{SmithGuptaViswanath:07ISIT}, the percolation-based
routing was applied under the exponential decay model. In our 3D
erasure network setup, we also introduce a routing protocol based
on the 3D percolation highway for both polynomial and exponential
decay models.

\subsection{Review on Percolation-Based Highway Routing}

The routing protocol based on the percolation
model~\cite{FranceschettiDouseTseThiran:07} is briefly reviewed.
When water {\it percolates} through one stone, the stone can be
modelled as a square grid, in which each edge is open with
probability $p$ and traversed by the water and is closed
otherwise. A wireless ad hoc network can also be modelled in a
similar fashion. A grid edge is open when there exists at least
one wireless transmitter(s) in the position corresponding to the
edge. Otherwise, it is closed. The open percolating paths can be
treated as a wireless backbone, named a {\it highway system}, that
conveys data packets across the network. The source nodes access
the highway system using single-hop transmission, and then the
information is carried via MH transmission using the highway
system across the network.

\begin{figure}[t!]
  \centering
  \includegraphics[width=0.57\textwidth]{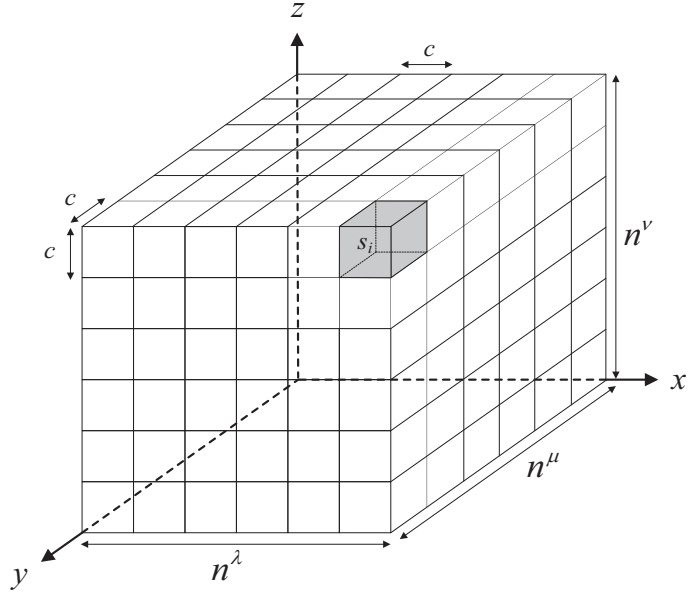}\\
  \caption{The tessellation of a cuboid network.}
  \label{Fig:Network}
\end{figure}

\subsection{Highway System}
Let us introduce a highway system in our 3D network. As
illustrated in Fig.~\ref{Fig:Network}, the cuboid is partitioned
into subcubes $s_i$ of constant side length $c>0$, independent of
$n$. Let $X(s_i)$ denote the number of nodes inside $s_i$, which
follows the binomial distribution with parameters $n$ and
$\frac{c^3}{n}$. For the situation in which $n$ is large, the
binomial distribution can be well-approximated by a Poisson
distribution with parameter $c^3$. Thus, similarly as
in~\cite{FranceschettiDouseTseThiran:07,ElGamalMammenPrabhakarShah:06,OzgurLevequeTse:07,ShinJeonDevroyeVuChungLeeTarokh:08},
the probability that a subcube contains at least one node is given
by
\begin{align*}
p :=\Pr\{X(s_i)\geq 1 \} \mathop  \approx \limits_{n \to \infty }
1 - e^{-c^3}.
\end{align*}
We call that a subcube is {\it open} if it contains at least one
node, and is {\it closed} otherwise. Then, the subcube is open
with probability $p$, or is closed with probability $1-p$.

\begin{figure}[t!]
  \centering
  \includegraphics[width=0.78\textwidth]{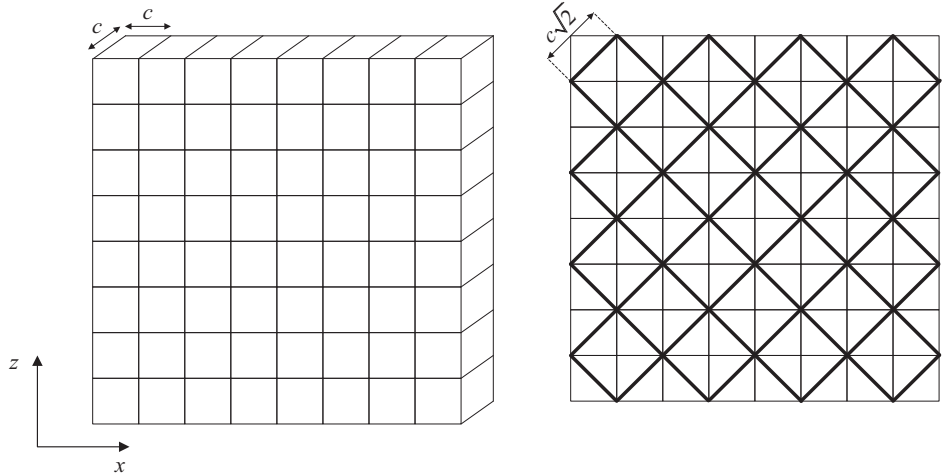}\\
  \caption{The construction of the percolation model along the $x$-$z$ plane.}
  \label{Fig:PercolationModel}
\end{figure}

Let us consider a cuboid of side length $n^{\lambda}\times c
\times n^{\nu}$, where a $y$-coordinate lies between $(i-1)c$ and
$ic$ for an integer $i$ (refer to the left-hand side of
Fig.~\ref{Fig:PercolationModel}). A vertical section $V_{xz}$ that
cuts $\frac{n^{\lambda}}{c}\times\frac{n^{\nu}}{c}$ subcubes in
the cuboid along the $x$-$z$ plane is depicted in the right-hand
side of Fig.~\ref{Fig:PercolationModel}. The section $V_{xz}$ is
tessellated into subsquares formed by projecting these subcubes
onto $V_{xz}$. We associate an edge to each square, traversing it
diagonally. An associated edge in the square lattice is called
{\it open} if the corresponding subcube contains at least one
node, and is {\it closed} otherwise. Each edge is open with
probability $p$, independently of each other. A path is formed by
connecting open edges, which are associated to subcubes that
contain at least one node.

\begin{figure}[t!]
  \centering
  \includegraphics[width=0.63\textwidth]{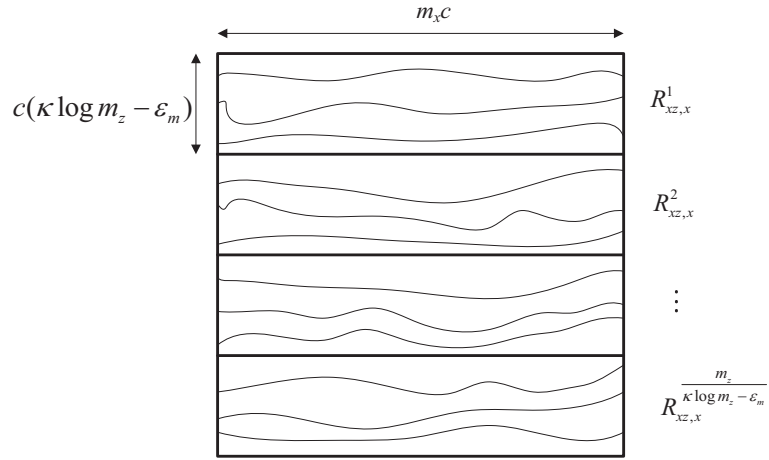}\\
  \caption{Many crossing paths that behave almost as straight lines from left to right.}
  \label{Fig:RectangularPartition}
\end{figure}

Let us denote the number of subcubes in $x$, $y$, and $z$
directions by $m_x$, $m_y$, and $m_z$, respectively. Then, it
follows that $m_x=\frac{n^{\lambda}}{c}$, $m_y=\frac{n^{\mu}}{c}$,
and $m_z=\frac{n^{\nu}}{c}$. Now, as illustrated in
Fig.~\ref{Fig:RectangularPartition}, let us partition $V_{xz}$
into rectangles $R_{xz,x}^{j}$ of sides $m_x c\times c(\kappa\log
m_z -\epsilon_m)$, where $\kappa>0$ is an arbitrary constant and
$\epsilon_m>0$ is chosen as the smallest value such that the
number of rectangles in $V_{xz}$, $\frac{m_z}{\kappa\log m_z
-\epsilon_m}$, is a positive integer. Similarly, $V_{xz}$ can be
partitioned into rectangles $R_{xz,z}^{j}$ of sides $c(\kappa\log
m_x -\epsilon_m)\times m_z c$, where $\epsilon_m>0$ is chosen as
the smallest value such that the number of rectangles in $V_{xz}$,
$\frac{m_x}{\kappa\log m_x -\epsilon_m}$, is a positive integer.
In the following lemma, it is shown that there are at least
$\delta\log m_z$ paths crossing each rectangle $R_{xz,x}^{j}$ from
left to right. It is also shown that there are at least
$\delta\log m_x$ paths crossing each rectangle $R_{xz,z}^j$ from
bottom to top.

\begin{lemma}\label{Lemma:MinHighwayPaths}
For all $\kappa>0$ and $\frac{5}{6}<p<1$ satisfying
$1+\frac{\lambda}{\nu}+\kappa\log(6(1-p))<0$, there exists a
$\delta>0$ such that
\begin{align}
\lim_{n\rightarrow\infty} \Pr\{N_{xz,x}\leq\delta\log m_z\}=0,
\nonumber
\end{align}
where $N_{xz,x}=\min_j C_{xz,x}^j$, $C_{xz,x}^j$ is the maximal
number of edge-disjoint left-to-right crossings of rectangle
$R_{xz,x}^j$, $j=1,\ldots,\frac{m_z}{\kappa\log m_z-\epsilon_m}$,
and $m_z=\frac{n^{\nu}}{c}$. For all $\kappa>0$ and
$\frac{5}{6}<p<1$ satisfying
$1+\frac{\nu}{\lambda}+\kappa\log(6(1-p))<0$, there exists a
$\delta>0$ such that
\begin{align}
\lim_{n\rightarrow\infty} \Pr\{N_{xz,z}\leq\delta\log m_x\}=0,
\nonumber
\end{align}
where $N_{xz,z}=\min_j C_{xz,z}^j$, $C_{xz,z}^j$ is the maximal
number of edge-disjoint bottom-to-top crossings of rectangle
$R_{xz,z}^j$, $j=1,\ldots,\frac{m_x}{\kappa\log m_x-\epsilon_m}$,
and $m_x=\frac{n^{\lambda}}{c}$.
\end{lemma}
\begin{IEEEproof}
Refer to Appendix~\ref{PF:NumHighwayPaths}.
\end{IEEEproof}

Hence, if we select $p$ sufficiently large, then the percolation
highways can be formed along $x$ and $z$ directions. The highways
in $y$ direction can be formed by either the section $V_{xy}$ or
$V_{yz}$. In what follows, it is assumed that the highways in $y$
direction are formed using horizontal paths in $V_{yz}$. Due to
the symmetry of the $x$, $y$, and $z$ coordinates, as in
Lemma~\ref{Lemma:MinHighwayPaths}, one can show that there also
exist highways along $y$ direction. This constitutes our highway
system in our cuboid network.

\subsection{Overall Procedure}
Similarly as in ~\cite{HuWangYangZhangXuGao:TC10}, the overall
procedure of the routing protocol in our 3D network is described.
Using the highway system in the network, the data packets can be
delivered towards any direction in 3D space. The routing protocol
under the 3D extended network consists of three phases and is
explained as follows:

\begin{figure}[t!]
  \centering
  \includegraphics[width=0.79\textwidth]{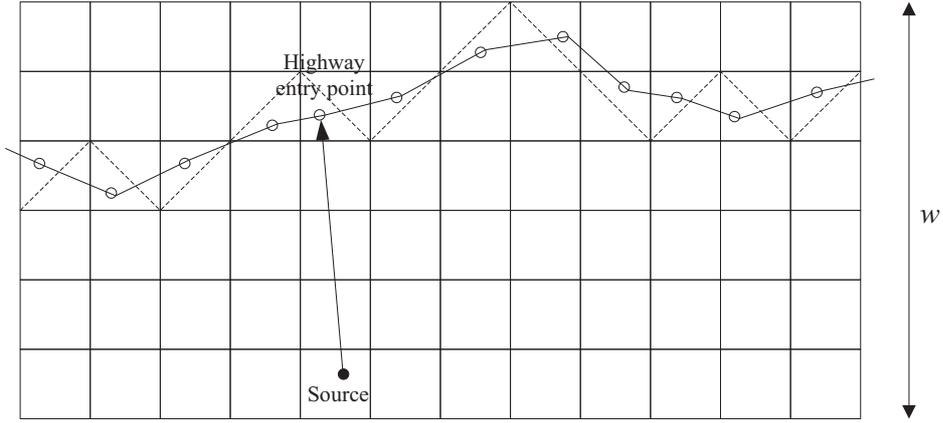}\\
  \caption{An illustration of the draining phase of the routing protocol.}
  \label{Fig:DrainingPhase}
\end{figure}

\begin{figure}[t!]
  \centering
  \includegraphics[width=0.87\textwidth]{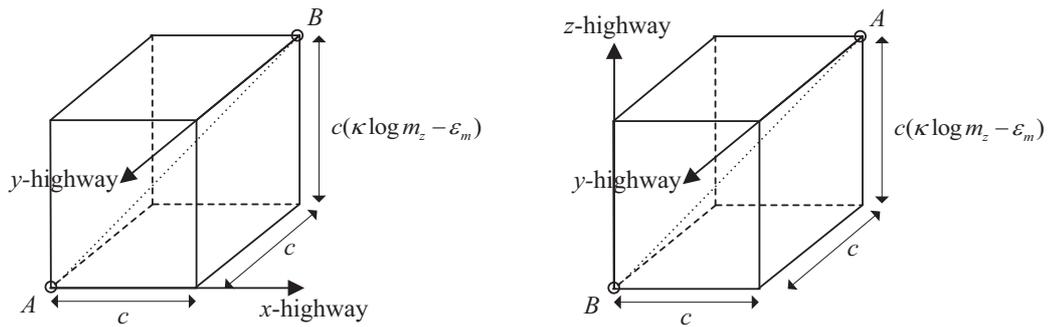}\\
  \caption{An illustration of the highway interchange phase of the routing protocol. In the left figure, the packets on the highway in $x$ direction are delivered to the highway in $y$ direction.
  In the right figure, the packets on the highway in $y$ direction are delivered to the highway in $z$ direction. The length of the dotted line from $A$ to $B$ is the farthest distance during each interchange step.}
  \label{Fig:InterchangePhase}
\end{figure}

\begin{itemize}
  \item (Draining phase) The section $V_{xz}$ perpendicular to $y$-axis is sliced into horizontal strips of constant width $w$, independent of $n$, where we choose $w>0$ appropriately such that there are at least as many paths as slices inside each rectangle and thus $w=\Theta(1)$. By imposing that nodes in the $i$th slice use the $i$th horizontal path for packet delivery, the traffic can evenly be distributed into all highways in a rectangle. Each source in the $i$th slice transmits its data packets to the entry point on the $i$th path in $x$ direction, where the entry point is the node on the path closest to the vertical line drawn from the source node, as illustrated in Fig.~\ref{Fig:DrainingPhase}.
  \item ($x$-highway phase) The packets are delivered along the highways in $x$ direction using MH routing until they reach the interchange point closest to the target highway in $y$ direction.
  \item (First highway interchange step) The packets are then delivered from the interchange point on the path in $x$ direction to another interchange point on a path in $y$ direction using single-hop, which is referred to as the first highway interchange step as illustrated in Fig.~\ref{Fig:InterchangePhase}.
  \item ($y$-highway phase) The packets are delivered along the highways in $y$ direction using MH routing until they reach the interchange point closest to the target highway in $z$ direction.
  \item (Second highway interchange step) The packets are then delivered from the interchange point on the path in $y$ direction to another interchange point on a path in $z$ direction using single-hop, which is referred to as the second highway interchange step as illustrated in Fig.~\ref{Fig:InterchangePhase}.
  \item ($z$-highway phase) The packets are delivered along the highways in $z$ direction using MH routing until they reach the exit point to the destination.
  \item (Delivery phase) The section $V_{xz}$ perpendicular to $y$-axis is sliced into vertical strips of constant width $w>0$ as in the draining phase. The packets are delivered from the exit point to the destination, where the exit point is the node on the path in $z$ direction closest to the horizontal line drawn from the destination.
\end{itemize}

\section{Achievability Results}\label{SEC:Achievability}

In this section, for two fundamental path-loss attenuation models
(i.e., exponential and polynomial decay models), the achievable
throughput scaling laws are analyzed in the 3D erasure network of
unit node density. For each decay model, the achievability proof
is provided according to the following steps. The transmission
rate $R(d)$ at which a node transmits to any destination located
within given distance $d>0$ is shown first. Using this rate, the
transmission rates achieved in the draining and delivery phases
are then derived. The transmission rate achievable along the 3D
highways is also shown. Based on the transmission rate in each
phase of the percolation-based highway routing, we finally present
the achievable throughput scaling.

We now establish several important binning lemmas that will be
conveniently used in the following two subsections to show our
achievability results.

\begin{lemma}\label{Lemma:NumNodesInSubcube}
If we partition the cuboid into an integer number
$l^3=\frac{n}{c^3}$ of subcubes $s_i$ of constant side length $c$,
then there are less than $\log l$ nodes in each subcube with high
probability (whp).
\end{lemma}
\begin{IEEEproof}
Let $A_n$ be the event that there is at least one subcube with
more than $\log l$ nodes. Since the number of nodes in each
subcube, $|s_i|$, is a Poisson random variable of parameter $c^3$,
by the union and Chernoff bounds, we have
\begin{align}
  \Pr\{A_n\}&\leq l^3 \Pr\{|s_i|>\log l\}
  \nonumber\\
  &\leq l^3 \frac{e^{-c^3}(c^3 e)^{\log l}}{\log l^{\log l}}
  \nonumber\\
  &= l^3 e^{-c^3}\left(\frac{c^3 e}{\log l}\right)^{\log l}
  \nonumber\\
&= e^{-c^3}\left(\frac{c^3 e^4}{\log l}\right)^{\log l},
  \nonumber
\end{align}
which approaches zero as $l$ tend to infinity. This completes the
proof of this lemma.
\end{IEEEproof}

\begin{lemma}\label{Lemma:NumNodesInCuboid}
If we partition the cuboid of side length $n^{\lambda}\times
c\times n^{\nu}$ into an integer number $\frac{n^{\nu}}{w}$ of
cuboids $C_i$ of side length $n^{\lambda}\times c\times w$, then
there are less than $2cwn^{\lambda}$ nodes in each cuboid $C_i$
whp.
\end{lemma}
\begin{IEEEproof}
Let $B_n$ be the event that there is at least one cuboid with more
than $2cwn^{\lambda}$ nodes. Since the number of nodes in each
cuboid, $|C_i|$, is a Poisson random variable of parameter
$cwn^{\lambda}$, by the union and Chernoff bounds, we have
\begin{align}
  \Pr\{B_n\}&\leq \frac{n^{\nu}}{w} \Pr\{|C_i|>2cwn^{\lambda}\}
  \nonumber\\
  &\leq \frac{n^{\nu}}{w}e^{-cwn^{\lambda}} \frac{(ecwn^{\lambda})^{2cwn^{\lambda}}}{(2cwn^{\lambda})^{2cwn^{\lambda}}}
  \nonumber\\
  &= \frac{n^{\nu}}{w}e^{-cwn^{\lambda}}\left(\frac{e}{2}\right)^{2cwn^{\lambda}}
  \nonumber\\
  &= \frac{n^{\nu}}{w}\left(\frac{\sqrt{e}}{2}\right)^{2cwn^{\lambda}}, \nonumber
\end{align}
which approaches zero as $n$ tends to infinity. This completes the
proof of this lemma.
\end{IEEEproof}

\subsection{Achievable Throughput Scaling Under the Exponential Decay Model}

Under the exponential decay model, as shown in
(\ref{Eq:ErasureProb-Exponential}), the probability of successful
transmission decays exponentially with a distance between two
nodes. The transmission rate $R(d)$ at which a node transmits to
any destination located within given distance $d>0$ is first
derived by using a TDMA operation, which is applied to avoid
causing any huge interference. In our work, the TDMA scheme that
finds the appropriate number of required time slots plays a key
role in obtaining the optimal scaling result.

\begin{lemma} \label{LEM:Rd}
Under the decay exponential model, there exists an $R(d)>0$ for
any integer $d>0$, such that, in each cube, there is a node that
can transmit at rate $R(d)$ to any destination located within
distance $d$ whp. Furthermore, as $d$ tends to infinity, we have
\begin{align}
  R(d) = \Omega(d^{-3}\gamma^{\sqrt{2}cd}), \nonumber
\end{align}
where $c>0$ is the side length of subcubes (see
Fig.~\ref{Fig:Network}).
\end{lemma}

\begin{IEEEproof}
Suppose that, based on the $t$-TDMA scheme, the time is divided
into a sequence of $t$ successive slots with $t=(k(d+1))^3$, where
$k>0$ is a constant (which will be specified later). Then, at each
time slot, multiple nodes that are at least distance
$(t^{1/3}-1)c$ away from each other transmit simultaneously, where
$c$ is the side length of each subcube $s_i$. A symbol transmitted
from one node can be decoded successfully if the symbol is not
erased and the other symbols from interfering nodes are all
erased. Since there are $(2i+1)^3-(2i-1)^3=2(12i^2+1)$ interfering
nodes whose distance from the intended receiver is given by
$(ki-1)(d+1)c$ in the $i$th layer, the probability $P_I$ that the
symbol from at least one of the simultaneously interfering nodes
is not erased is given by
\begin{align}\label{Eq:PiBound}
    P_I&\leq \sum_{i=1}^{\infty}2(12i^2+1)\gamma^{(ki-1)(d+1)c}
    \nonumber\\
    &\leq\sum_{i=1}^{\infty}2(12i^2+i^2)\gamma^{(ki-i)(d+1)c}
    \nonumber\\
    &=26\sum_{i=1}^{\infty}i^2\gamma^{(k-1)c(d+1)i}
    \nonumber\\
    &=\frac{26x(1+x)}{(1-x)^3}\leq 1,
\end{align}
where
\begin{align}
x=\gamma^{(k-1)c(d+1)}. \label{Eq:x_gamma}
\end{align}
Using (\ref{Eq:PiBound}), we have
\begin{align}
    x^3 + 23x^2 + 29x - 1 \leq 0. \nonumber
\end{align}
Hence, from (\ref{Eq:PiBound}) and (\ref{Eq:x_gamma}), it follows
that the probability $P_I$ is less than one if
\begin{align}
    k\geq 1+\frac{\log_2 y}{c(d+1)\log_2 \gamma}, \nonumber
\end{align}
where $y>0$ is the greatest root of the equation $x^3 + 23x^2 +
29x - 1=0$. From the fact that the distance between the
transmitter and the receiver is at most $c\sqrt{2(d+1)^2+1}$, the
probability that a symbol from the intended transmitter is not
erased is given by
\begin{align}
\gamma^{c\sqrt{2(d+1)^2+1}}.\nonumber
\end{align}
Finally, using the $t$-TDMA with $t=\Theta(d^3)$ time slots leads
to the achievable transmission rate
$\Omega(d^{-3}\gamma^{\sqrt{2}cd})$ in each cube, which completes
the proof of this lemma.
\end{IEEEproof}

Now, the achievable transmission rate in the draining and delivery
phases of the routing protocol is derived in the following lemma.

\begin{lemma}\label{Lemma:RateDrainingDelivery-Exponential}
Suppose a 3D erasure network of size length $n^\lambda \times
n^\mu \times n^\nu$ under the exponential decay model. In the
draining phase, every node in a cube can achieve a transmission
rate of $\Omega\left((\log
n)^{-4}n^{-\frac{\sqrt{2}c\kappa\nu}{d^*}} \right)$ to a certain
node on the highway system whp, where $d^*>0$ is the critical
distance such that $\gamma^d=e^{-d/d^*}$. In the delivery phase,
every destination node can successfully receive information from
the highway with rate $\Omega\left((\log
n)^{-4}n^{-\frac{\sqrt{2}c\kappa\lambda}{d^*}} \right)$ whp.
\end{lemma}

\begin{IEEEproof}
Let us first focus on analyzing the transmission rate in the
draining phase. We note that the distance between sources and
entry points is never greater than $c(\kappa\log m_z+\sqrt{2})$
from Lemma~\ref{Lemma:MinHighwayPaths} and the triangle
inequality. From Lemma~\ref{LEM:Rd}, we obtain that one node per
cube can communicate with its entry point at rate
\begin{align}
    R(\kappa\log m_z + \sqrt{2})
    &=R\left(\kappa\log\frac{n^{\nu}}{c}+\sqrt{2}\right), \nonumber
\end{align}
which is further lower-bounded by
\begin{align}
    \Omega\left(\frac{\gamma^{\sqrt{2}c\kappa\log\frac{n^{\nu}}{c}}}{\left(\kappa\log\frac{n^{\nu}}{c}\right)^3}
    \right)
    &=\Omega\left(\frac{e^{-\frac{\sqrt{2}c\kappa}{d^*}\log\frac{n^{\nu}}{c}}}{\left(\kappa\log\frac{n^{\nu}}{c}\right)^3}
    \right)
    \nonumber\\
    &=\Omega\left(\frac{n^{-\frac{\sqrt{2}c\kappa\nu}{d^*}}}{\left(\log n\right)^3}
    \right). \nonumber
\end{align}
Now we note that, as there are possibly many nodes in the
subcubes, the nodes have to share their assigned bandwidth. Hence,
using Lemma~\ref{Lemma:NumNodesInSubcube}, we finally obtain that
the transmission rate of each node in the draining phase of our
protocol is at least $R(d)/\log \left(\frac{n^{1/3}}{c}\right)$.
The achievable transmission rate in the delivery phase can be
derived in a similar way, which completes the proof of this lemma.
\end{IEEEproof}

Next, we establish the two lemmas below, which show the
transmission rate along the highways in three Cartesian directions
and the transmission rate during the interchange steps.

\begin{lemma}\label{Lemma:RateAlongHighways-Exponential}
Suppose a 3D erasure network of size length $n^\lambda \times
n^\mu \times n^\nu$ under the exponential decay model. The nodes
along the highways in $x$, $y$, and $z$ directions can achieve
per-node transmission rate of $\Omega\left(n^{-\lambda}\right)$,
$\Omega\left(n^{-\mu}\right)$, $\Omega\left(n^{-\nu}\right)$,
respectively, whp.
\end{lemma}

\begin{IEEEproof}
We divide the whole highway into three routes according to $x$,
$y$, and $z$ directions. Let us start by considering the traffic
flow in $x$ direction. Let a node be sitting on the $i$th
$x$-directional highway and compute the traffic that goes through
it. Notice that, at most, the node will relay all the traffic
generated in the $i$th cuboid of side length $n^{\lambda}\times
c\times w$. According to Lemma~\ref{Lemma:NumNodesInCuboid}, a
node on the $x$-directional highway must relay the traffic for at
most $2cwn^{\lambda}$ nodes. As the maximal distance between hops
is constant, i.e., $\Theta(1)$, by Lemma~\ref{LEM:Rd}, an
achievable transmission rate along the highways is
$\Omega(n^{-\lambda})$ whp.

The problem of the traffic flow in $y$ and $z$ directions is the
dual of the previous one. Thus, we can apply the same argument to
compute the transmission rate achieved by each node on the $y$-
and $z$-directional highways. This completes the proof of this
lemma.
\end{IEEEproof}

\begin{lemma}\label{Lemma:RateInterchange-Exponential}
Suppose a 3D erasure network of size length $n^\lambda \times
n^\mu \times n^\nu$ under the exponential decay model. During the
first interchange step, every node on the highway in $x$ direction
can achieve a transmission rate of $\Omega\left((\log
n)^{-4}n^{-\frac{\sqrt{2}c\kappa\nu}{d^*}} \right)$ to a certain
node on the highway in $y$ direction whp, where $d^*>0$ is the
critical distance such that $\gamma^d=e^{-d/d^*}$. During the
second interchange step, every node on the highway in $y$
direction can achieve a transmission rate of $\Omega\left((\log
n)^{-4}n^{-\frac{\sqrt{2}c\kappa\nu}{d^*}} \right)$ to a certain
node on the highway in $z$ direction whp.
\end{lemma}
\begin{IEEEproof} Let us first focus on the case
where the packet is delivered from one interchange point on the
highway in $x$ direction to another interchange point on the
highway in $y$ direction (i.e., the fist interchange step). We
note that the distance between interchange points is never greater
than $c(\kappa\log m_z+ \sqrt{2})$ from
Lemma~\ref{Lemma:MinHighwayPaths} and the triangle inequality, as
illustrated in Fig.~\ref{Fig:InterchangePhase}. From
Lemma~\ref{LEM:Rd}, we obtain that one node per cube can
communicate with its entry point at rate
\begin{align}
    R(\kappa\log m_z + 1)
    &=R\left(\kappa\log\frac{n^{\nu}}{c}+1\right) \nonumber\\
    &=\Omega\left(\frac{\gamma^{\sqrt{2}c\kappa\log\frac{n^{\nu}}{c}}}{\left(\kappa\log\frac{n^{\nu}}{c}\right)^3}
    \right)\nonumber\\
    &=\Omega\left(\frac{e^{-\frac{\sqrt{2}c\kappa}{d^*}\log\frac{n^{\nu}}{c}}}{\left(\kappa\log\frac{n^{\nu}}{c}\right)^3}
    \right)
    \nonumber\\
    &=\Omega\left(\frac{n^{-\frac{\sqrt{2}c\kappa\nu}{d^*}}}{\left(\nu\log n\right)^3}
    \right). \nonumber
\end{align}
From the fact that there are possibly many nodes in the subcubes,
using Lemma~\ref{Lemma:NumNodesInSubcube}, we finally obtain that
the transmission rate of each node in the first exchange step of
the 3D highway phase is at least $R(d)/\log (\frac{n^{1/3}}{c})$.
The achievable transmission rate in the second interchange step
can be derived in a similar way. This completes the proof of this
lemma.
\end{IEEEproof}

Using the aforementioned lemmas, we finally present the achievable
throughput scaling for the exponential decay model in the 3D
erasure network in the following theorem.

\begin{theorem}\label{Theorem:Achievability-Exponential}
Suppose a 3D erasure network of size length $n^\lambda \times
n^\mu \times n^\nu$ with unit node density under the exponential
decay model, where the probability of successful transmission
decays exponentially as in (\ref{Eq:ErasureProb-Exponential}).
Then, the aggregate throughput $T_n$ is lower-bounded by
\begin{align}
    T_n=\Omega(n^{\min\{1-\lambda,1-\mu,1-\nu\}}). \nonumber
\end{align}
\end{theorem}

\begin{IEEEproof}
From
Lemmas~\ref{Lemma:RateDrainingDelivery-Exponential}--\ref{Lemma:RateInterchange-Exponential},
the overall per-node transmission rate is limited by the highway
phase only if
\begin{align}\label{Eq:Condition1-Exponential}
    \frac{\sqrt{2}c\kappa\max\{\lambda,\nu\}}{d^*}<\min\{\lambda,\mu,\nu\},
\end{align}
where $c>0$ is the side length of subcubes (refer to
Fig.~\ref{Fig:Network}). We can choose $c$ and $\kappa$ such that
the highways are formed as in Lemma~\ref{Lemma:MinHighwayPaths}
and (\ref{Eq:Condition1-Exponential}) is satisfied. Therefore, the
aggregate throughput is given by
$\Omega\left(n^{\min\{1-\lambda,1-\mu,1-\nu\}}\right)$ since the
minimum rate along the highways is
$\Omega\left(n^{-\max\{\lambda,\mu,\nu\}}\right)$, which completes
the proof of Theorem~\ref{Theorem:Achievability-Exponential}.
\end{IEEEproof}

From the achievability result for the exponential decay model, the
following interesting observations can be made.

\begin{remark}
Under the exponential decay model, if we use the nearest-neighbor
MH routing~\cite{GuptaKumar:00} instead of the percolation-based
highway routing, then we cannot achieve a constant throughput
scaling $\Omega(1)$ for each hop. This is because, under the
nearest-neighbor MH routing protocol, the probability of
successful transmission decays exponentially with the distance
while per-hop distance increases in logarithmic scale.
\end{remark}

\begin{remark}
For a cubic network where $\lambda=\mu=\nu=\frac{1}{3}$, it is
shown under the exponential decay model that the achievable
throughput scaling $\Omega(n^{2/3})$ is higher than the throughput
scaling $\Omega(\sqrt{n})$ in 2D square erasure
networks~\cite{SmithGuptaViswanath:07ISIT}. As in the 2D network
topology, a bottleneck of data transmission in the 3D network is
the transmission along the highways. Since there are additional
orthogonal directions in 3D space compared to the 2D network case,
the burden of packet forwarding can be significantly reduced in 3D
space, thereby resulting in a higher performance on the
throughput. In other words, compared to 2D space, more {\em
geographic diversity} can be exploited via 3D geolocation while
generating more simultaneous end-to-end percolation highways is
possible.
\end{remark}

\begin{remark} Let us consider a 3D network
configuration whose height is constant, i.e. does not scale with
$n$, by setting $\lambda=\mu=\frac{1}{2}$ and $\nu=0$. From
Theorem~\ref{Theorem:Achievability-Exponential}, the aggregate
throughput is given by $T_n=\Omega(\sqrt{n})$, which is
essentially the same as the 2D network case. Thus, our result is
general in the sense that it includes the existing achievability
result obtained from 2D space.
\end{remark}

\subsection{Achievable Throughput Scaling Under the Polynomial Decay Model}

Besides the exponential decay model, in which the probability of
successful transmission decays exponentially with a distance
between two nodes, another fundamental path-loss attenuation model
is the polynomial decay model, where the probability of successful
transmission decays polynomially with a distance between two
nodes, as shown in (\ref{Eq:ErasureProb-Polynomial}). We first
derive the transmission rate $R(d)$ for a given distance $d$ based
on the TDMA operation.

\begin{lemma} \label{LEM:Rd2}
Under the polynomial decay model, there exists an $R(d)>0$ for any
integer $d>0$, such that, in each cube, there is a node that can
transmit at rate $R(d)$ to any destination located within distance
$d$ whp. Furthermore, as $d$ tends to infinity, we have
\begin{align}
  R(d) = \Omega(d^{-\alpha-3}), \nonumber
\end{align}
where $\alpha>3$.
\end{lemma}

\begin{IEEEproof}
Similarly as in the exponential decay model, suppose that the time
is divided into a sequence of $t$ successive slots with
$t=(k(d+1))^3$. According to the same argument as the proof of
Lemma~\ref{LEM:Rd}, we obtain that there are $2(12i^2+1)$
interfering nodes whose distance from the intended receiver is
given by $(ki-1)(d+1)c$ in the $i$th layer, where $c>0$ is the
side length of subcubes (refer to Fig.~\ref{Fig:Network}). Thus,
the probability $P_I$ that the symbol from at least one of the
simultaneously interfering nodes is not erased is given by
\begin{align}
    &P_I\nonumber\\
    &\leq \sum_{i=1}^{\infty}
    \frac{2(12i^2+1)}{((ki-1)(d+1)c)^\alpha}
    \nonumber\\
    &=\frac{1}{(c(d+1))^{\alpha}}\sum_{i=1}^{\infty} \frac{2(12i^2+1)}{(ki-1)^\alpha}
    \nonumber\\
    &=\frac{1}{(c(d+1))^{\alpha}}
    \nonumber\\
    &~~~\cdot\left(\frac{2(12\cdot 1^2+1)}{(k\cdot 1 -1)^\alpha} \!+\! \frac{2(12\cdot 2^2+1)}{(k\cdot 2 -1)^\alpha}
    + \frac{2(12\cdot 3^2+1)}{(k\cdot 3 -1)^\alpha} \!+\! \cdots\!\right)
    \nonumber\\
    &\leq \frac{1}{(c(d+1))^{\alpha}}\cdot
    \nonumber\\
    &~~\left(\frac{2(12\cdot 1^2+1)}{(k -1)^\alpha} + \frac{2(12\cdot 2^2+1)}{k^{\alpha}}
    + \frac{2(12\cdot 3^2+1)}{(2k)^\alpha} + \cdots\right)
    \nonumber\\
    &=\frac{2(12\cdot 1^2+1)}{(c(d+1))^{\alpha}(k -1)^\alpha}
    + \frac{1}{(c(d+1))^{\alpha}}\sum_{i=1}^{\infty}\frac{2(12(i+1)^2+1)}{(ki)^\alpha}
    \nonumber
    \end{align}
    \begin{align}
    &=\frac{26}{((k-1)c(d+1))^{\alpha}}
    \nonumber\\ &~~
    + \frac{2}{(kc(d+1))^{\alpha}}\sum_{i=1}^{\infty}\left(
    \frac{12}{i^{\alpha-2}}+\frac{24}{i^{\alpha-1}}+\frac{13}{i^{\alpha}}
    \right)
    \nonumber\\
    &\leq \frac{26}{((k-1)c(d+1))^{\alpha}}
    \nonumber\\ &~~
    + \frac{2}{((k-1)c(d+1))^{\alpha}}\sum_{i=1}^{\infty}\left(
    \frac{12}{i^{\alpha-2}}+\frac{24}{i^{\alpha-1}}+\frac{13}{i^{\alpha}}
    \right)
    \nonumber\\
    &=\frac{2}{((k-1)c(d+1))^{\alpha}}\left( 13 + \sum_{i=1}^{\infty}\left(
    \frac{12}{i^{\alpha-2}}+\frac{24}{i^{\alpha-1}}+\frac{13}{i^{\alpha}}
    \right)
    \right), \nonumber
\end{align}
where $k>1$ and the sum in the last equality converges when
$\alpha>3$. Let $K_\alpha$ denote the term
$\sum_{i=1}^{\infty}\left(\frac{12}{i^{\alpha-2}}+\frac{24}{i^{\alpha-1}}+\frac{13}{i^{\alpha}}\right)$.
Then, given the value of $\alpha>3$, the probability $P_I$ is
shown to be less than one when the value of $k$ is set to
\begin{align}
    k > 1 + \frac{(2(13+K_{\alpha}))^{1/\alpha}}{c(d+1)}. \nonumber
\end{align}
Hence, it follows that
\begin{align}
    t>\left(d+1 + \frac{(2(13+K_{\alpha}))^{1/\alpha}}{c} \nonumber
    \right)^3.
\end{align}
Due to the fact that the distance between the transmitter and the
receiver is at most $c\sqrt{2(d+1)^2+1}$, the probability that a
transmitted symbol is not erased is given by
$\left(c\sqrt{2(d+1)^2+1}\right)^{-\alpha}$.
Finally, using the $t$-TDMA with $t=\Theta(d^3)$ time slots, we
have that the transmission rate available in each cube is
$\Omega(d^{-\alpha-3})$, which completes the proof of this lemma.
\end{IEEEproof}

In the following lemma, the achievable transmission rate in the
draining and delivery phases of the routing protocol is derived
for the polynomial decay model.

\begin{lemma}\label{Lemma:RateDrainingDelivery-Polynomial}
Suppose a 3D erasure network of size length $n^\lambda \times
n^\mu \times n^\nu$ under the polynomial decay model. In the
draining phase, when $\alpha>3$, every node in a cube can achieve
a transmission rate of $\Omega\left((\log n)^{-4-\alpha}\right)$
to a certain node on the highway system whp. In the delivery
phase, when $\alpha>3$, every destination node can successfully
receive information from the highway with rate $\Omega\left((\log
n)^{-4-\alpha}\right)$ whp.
\end{lemma}

\begin{IEEEproof}
Let us focus on analyzing the transmission rate in the draining
phase. From Lemma~\ref{LEM:Rd2} and the fact that the distance
between sources and entry points is never greater than
$c(\kappa\log m_z+\sqrt{2})$, one node per cube can communicate
with its entry point at rate
\begin{align}
    R(\kappa\log m_z +\sqrt{2})&=R\left(\kappa\log\frac{n^{\nu}}{\sqrt{2}c}+\sqrt{2}\right)
    \nonumber\\
    &=\Omega\left(\frac{1}{\left(\kappa\log\frac{n^{\nu}}{\sqrt{2}c}\right)^{3+\alpha}}
    \right)
    \nonumber\\
    &=\Omega\left(\frac{1}{\left(\log n\right)^{3+\alpha}}
    \right). \nonumber
\end{align}
Hence, using Lemma~\ref{Lemma:NumNodesInSubcube}, we conclude that
the transmission rate of each node in the draining phase of our
protocol is at least $R(d)/\log (\frac{n^{1/3}}{c})$. The
achievable transmission rate in the delivery phase can be
similarly derived, which completes the proof of this lemma.
\end{IEEEproof}

Next, we establish the following two lemmas, which show the
transmission rate along the highways in three Cartesian directions
and the transmission rate during the interchange steps.

\begin{lemma}\label{Lemma:RateAlongHighways-Polynomial}
Suppose a 3D erasure network of size length $n^\lambda \times
n^\mu \times n^\nu$ under the polynomial decay model. The nodes
along the highways in $x$, $y$, and $z$ directions can achieve
per-node transmission rate of $\Omega\left(n^{-\lambda}\right)$,
$\Omega\left(n^{-\mu}\right)$, $\Omega\left(n^{-\nu}\right)$,
respectively, whp.
\end{lemma}

The proof of this lemma essentially follows the same line as that
of Lemma~\ref{Lemma:RateAlongHighways-Exponential} and thus is
omitted for brevity.

\begin{lemma}\label{Lemma:RateInterchange-Polynomial}
Suppose a 3D erasure network of size length $n^\lambda \times
n^\mu \times n^\nu$ under the polynomial decay model. During each
interchange step, when $\alpha>3$, every node on the highway in
one direction ($x$ or $y$ direction) can achieve a transmission
rate of $\Omega\left((\log n)^{-4-\alpha}\right)$ to a certain
node on the highway in other direction ($y$ or $z$ direction) whp.
\end{lemma}
\begin{IEEEproof} Let us focus on the fist
interchange step. The distance between these two interchange
points is never greater than $c(\kappa\log m_z+ \sqrt{2})$ from
Lemma~\ref{Lemma:MinHighwayPaths} and the triangle inequality, as
illustrated in Fig.~\ref{Fig:InterchangePhase}. From
Lemma~\ref{LEM:Rd}, one node per cube can communicate with its
entry point at rate
\begin{align}
    R(\kappa\log m_z +\sqrt{2})&=R\left(\kappa\log\frac{n^{\nu}}{\sqrt{2}c}+\sqrt{2}\right)
    \nonumber\\
    &=\Omega\left(\frac{1}{\left(\kappa\log\frac{n^{\nu}}{\sqrt{2}c}\right)^{3+\alpha}}
    \right)
    \nonumber\\
    &=\Omega\left(\frac{1}{\left(\log n\right)^{3+\alpha}}
    \right). \nonumber
\end{align}
As in the proof of Lemma~\ref{Lemma:RateInterchange-Exponential},
the transmission rate of each node in the first exchange step of
the 3D highway phase is lower-bounded by $R(d)/\log
(\frac{n^{1/3}}{c})$. The achievable transmission rate in the
second exchange step can be derived in a similar way, which
completes the proof of this lemma.
\end{IEEEproof}

Finally, we are ready to analyze the achievable throughput scaling
for the polynomial decay model in the 3D erasure network.

\begin{theorem}\label{Theorem:Achievability-Polynomial}
Suppose a 3D erasure network of size length $n^\lambda \times
n^\mu \times n^\nu$ with unit node density under the polynomial
decay model, where the probability of successful transmission
decays polynomially as in (\ref{Eq:ErasureProb-Polynomial}). Then,
when $\alpha>3$, the aggregate throughput $T_n$ is lower-bounded
by
\begin{align}
    T_n=\Omega\left(n^{\min\{1-\lambda,1-\mu,1-\nu\}}\right). \nonumber
\end{align}
\end{theorem}

\begin{IEEEproof}
From
Lemmas~\ref{Lemma:RateDrainingDelivery-Polynomial}--\ref{Lemma:RateInterchange-Polynomial},
the overall per-node transmission rate is limited by the highway
phase. As in Lemma~\ref{Lemma:MinHighwayPaths}, one can determine
$c$ and $\kappa$ such that the highways are formed. Therefore, the
aggregate throughput is given by
$\Omega\left(n^{\min\{1-\lambda,1-\mu,1-\nu\}}\right)$ since the
minimum rate along the highways is
$\Omega\left(n^{-\max\{\lambda,\mu,\nu\}}\right)$, which completes
the proof of Theorem~\ref{Theorem:Achievability-Polynomial}.
\end{IEEEproof}

From the achievability result for the polynomial decay model, the
following interesting observations can be made.

\begin{remark}
Unlike the case of the exponential decay model, it can be shown
that, under the polynomial decay model, using the nearest-neighbor
MH routing leads to the same throughput scaling behavior as that
of the percolation-based highway routing within a polylogarithmic
factor.
\end{remark}

\begin{remark}
Let us recall the scaling results for wireless Gaussian channels,
in which the received signal power decays polynomially with
distance. When a cubic network is assumed, i.e.,
$\lambda=\mu=\nu=\frac{1}{3}$, under the polynomial decay model in
the 3D erasure network, the same achievable throughput scaling is
achieved as in the 3D Gaussian network
scenario~\cite{HuWangYangZhangXuGao:TC10} using a routing protocol
based on the percolation theory. We remark that, in 2D networks,
using a percolation-based highway routing in the erasure network
model~\cite{SmithGuptaViswanath:07} achieves the same throughput
scaling law as that of the Gaussian network model based on the
percolation theory~\cite{FranceschettiDouseTseThiran:07}, which is
consistent with the achievability result in 3D space.
\end{remark}

\section{Cut-Set Upper Bounds}\label{SEC:UpperBound}

In this section, to verify the order optimality of our
achievability in Section~\ref{SEC:Achievability},
information-theoretic cut-set upper bounds~\cite{CoverThomas:91}
are derived for a 3D erasure network of unit node density. We
consider three cut planes $L_x$, $L_y$, and $L_z$ that are
perpendicular to $x$-, $y$-, and $z$-axes, respectively. Upper
bounds under the cut planes $L_x$, $L_y$, and $L_z$ are denoted by
${T}_{n,x}$, ${T}_{n,y}$, and ${T}_{n,z}$. By the max-flow min-cut
theorem, the aggregate throughput $T_n$ is then bounded by
\begin{align}
  {T}_n \le \min\{{T}_{n,x},{T}_{n,y},{T}_{n,z}\}. \nonumber
\end{align}
In what follows, we focus only on an analysis obtained from the
cut plane $L_x$. The other results from $L_y$ and $L_z$ can be
similarly derived.

\begin{figure}[t!]
  \centering
  \includegraphics[width=0.58\textwidth]{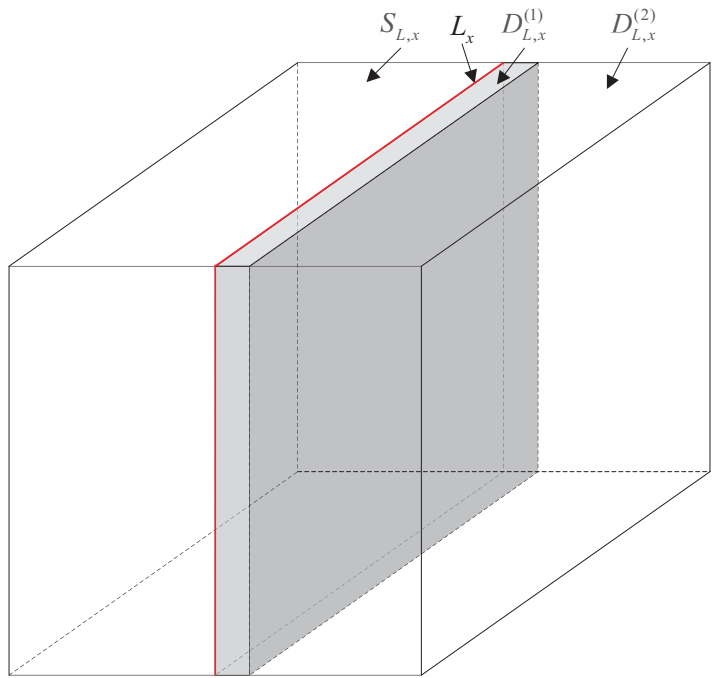}\\
  \caption{The sources and the partition of destinations with cut $L_x$.}
  \label{Fig:CutSet}
\end{figure}

Consider a cut plane $L_x$ that divides the 3D network into two
equal halves, each of which contains $n/2$ nodes, as illustrated
in Fig.~\ref{Fig:CutSet}. The set of destinations, $D_{L,x}$, is
further partitioned into two groups $D_{L,x}^{(1)}$ and
$D_{L,x}^{(2)}$ according to their locations. Here,
$D_{L,x}^{(1)}$ denote the sets of destinations located on the
cuboid with width one immediately to the right of the cut plane,
and $D_{L,x}^{(2)}$ is given by $D_{L,x}\setminus D_{L,x}^{(1)}$.
Note that the set $D_{L,x}^{(1)}$ of destinations located very
close to the cut plane $L_x$ are taken into account separately
since, otherwise, their contribution to the aggregate throughput
will be excessive, resulting in a loose bound. We start from the
following lemma, in which the cut-set bound for erasure networks
is characterized assuming no interference, leading to an upper
bound on the performance.

\begin{lemma}[\cite{DanaGowaikarHassibi:06}]
For an erasure network divided into two sets $S_{L,x}$ and
$D_{L,x}$, the cut-set bound on the aggregate throughput $T_n$ is
given by
\begin{align}
T_n \le \sum_{i\in S_{L,x}} \left(1-\prod_{k\in D_{L,x}}
\epsilon_{ki}\right), \label{EQ:upper}
\end{align}
where $\epsilon_{ki}$ is the erasure probability between source
$i\in S_{L,x}$ and destination $k\in D_{L,x}$.
\end{lemma}

Using the characteristics of random node distribution establishes
the following binning lemma.

\begin{lemma}\label{Lemma:NumNodesInsideCube}
Let the 3D network volume be divided into $n$ cubes of unit
volume. Then, there are less than $\log n$ nodes inside all cubes
whp.
\end{lemma}

\begin{IEEEproof}
This lemma can be proved by slightly modifying the proof
of~\cite[Lemma 1]{FranceschettiDouseTseThiran:07}.
\end{IEEEproof}

\begin{figure}[t!]
  \centering
  \includegraphics[width=0.5\textwidth]{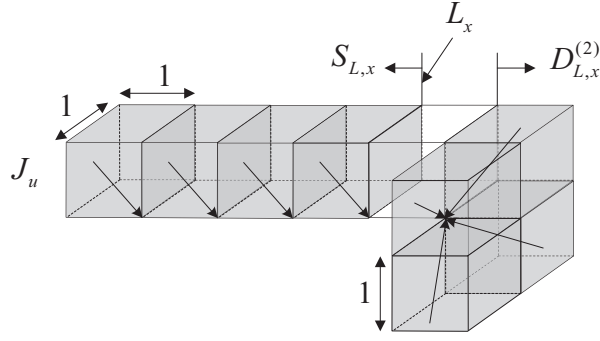}\\
  \caption{The displacement of the nodes inside the cubes to vertices, indicated by arrows.}
  \label{Fig:DisplacementNodes}
\end{figure}

 The cut-set upper
bound on the aggregate throughput $T_n$ is given by
\begin{align}
    T_n \leq T_n^{(1)} + T_n^{(2)}, \nonumber
\end{align}
where $T_n^{(1)}$ and $T_n^{(2)}$ denote the throughputs from the
set of sources, $S_{L,x}$, to the sets of corresponding
destinations, $D_{L,x}^{(1)}$ and $D_{L,x}^{(2)}$, respectively.
The contribution to $T_n^{(1)}$ from nodes in $D_{L,x}^{(1)}$ is
no greater than one for each node. Since there are no more than
$n^{\mu+\nu}\log n$ nodes in $D_{L,x}^{(1)}$ from
Lemma~\ref{Lemma:NumNodesInsideCube}, the throughput for the set
$D_{L,x}^{(1)}$ is upper-bounded by
\begin{align}
T_n^{(1)}=O(n^{\mu+\nu}\log n). \nonumber
\end{align}
Hence, from (\ref{EQ:upper}), an upper bound on $T_n$ is given by
\begin{align}
    T_n&\leq a_0n^{\mu+\nu}\log n+\sum_{i\in S_{L,x}}\left(1-\prod_{k\in D_{L,x}^{(2)}}\epsilon_{ki}\right)
    \nonumber\\
    &\leq a_0 n^{\mu+\nu}\log n+\sum_{i\in S_{L,x}}\sum_{k\in
    D_{L,x}^{(2)}}(1-\epsilon_{ki}), \nonumber
\end{align}
where $a_0>0$ is some constant independent of $n$. In order to
derive an upper bound on $T_n^{(2)}$, we would like to consider
the network transformation resulting in a {\em regular} network
with at most $\log n$ nodes in each subcube of unit volume,
similarly as
in~\cite{OzgurLevequeTse:07,ShinJeonDevroyeVuChungLeeTarokh:08}.
In this case, we can construct the resulting regular network in
which two neighboring nodes are regularly 1 unit of distance apart
from each other. Let us divide the left half of the network into
$n^{\mu+\nu}$ cuboids of side length $\frac{1}{2}n^{\lambda}\times
1 \times 1$. Let $J_u$ denote the $u$th cuboid of $S_{L,x}$, i.e.,
$S_{L,x}=\bigcup_{u=1}^{n^{\mu+\nu}}J_u$. Then, $T_n^{(2)}$ is
bounded by
\begin{align}
  T_n^{(2)}\leq \sum_{u=1}^{n^{\mu+\nu}} \sum_{i\in J_u}\sum_{k\in
  D_{L,x}^{(2)}}(1-\epsilon_{ki}). \label{Eq:Tn2}
\end{align}
As depicted in Fig.~\ref{Fig:DisplacementNodes}, as we move the
nodes that lie in each cube of $J_m$ together with the nodes in
the cubes of $D_{L,x}^{(2)}$ onto the vertex indicated by the
arrows, $T_n^{(2)}$ increases since this node displacement leads
to a decrement of the Euclidean distance between the associated
nodes. Since there are no more than $\log n$ nodes in each unit
cube, the modification results in a regular network with at most
$\log n$ nodes at each cube vertex on the left and at most $4\log
n$ nodes at each cube vertex on the right. Thus, the throughput
$T_n^{(2)}$ is less than the quantity achieved for a regular
network with $\log n$ nodes at each left-hand side vertex and
$4\log n$ nodes at each right-hand side vertex. In the following
two subsections, we derive upper bounds on the aggregate capacity
according to the two path-loss attenuation models.

\subsection{Upper Bound Under the Exponential Decay Model}

In this subsection, we present a cut-set upper bound on the
capacity for the exponential decay model by considering the
network transformation to a regular network.

\begin{theorem} \label{THM:upper_exp}
Suppose a 3D erasure network of size length $n^\lambda \times
n^\mu \times n^\nu$ with unit node density under the exponential
decay model, where the probability of successful transmission
decays exponentially as in (\ref{Eq:ErasureProb-Exponential}).
Then, the aggregate throughput is upper-bounded by
\begin{align}
  T_n=O(n^{\min\{1-\lambda,1-\mu,1-\nu\}}(\log n)^2). \nonumber
\end{align}
\end{theorem}

\begin{IEEEproof}
We start by assuming a regular network having one node at each
unit cube vertex. We first consider the cut plane $L_x$. Suppose
that the left-hand side nodes are located at positions
$(-i_l+1,j_l,k_l)$ and those on the right-hand side are located at
positions $(i_r,j_r,k_r)$. Substituting
(\ref{Eq:ErasureProb-Exponential}) into (\ref{Eq:Tn2}),
$T_n^{(2)}$ is bounded by
\begin{align}
    &T_n^{(2)}\nonumber\\
    &\leq \sum_{i\in S_L}\sum_{k\in D_L}\gamma^{d_{ki}}
    \nonumber\\
    &\leq\!\!\! \sum_{i_l=1,i_r=1}^{n^{\lambda}/2}\! \sum_{j_l=1,j_r=1}^{n^{\mu}}
    \!\sum_{k_l=1,k_r=1}^{n^{\nu}}\!\!\!
    \gamma^{((i_l+i_r-1)^2+(j_l-j_r)^2+(k_l-k_r)^2)^{1/2}}
    \nonumber\\
    &= \sum_{i_r=1}^{n^{\lambda}/2} \sum_{j_r=1}^{n^{\mu}} \sum_{k_r=1}^{n^{\nu}}
    \nonumber\\&~~~
    \left( \sum_{i_l=1}^{n^{\lambda}/2} \sum_{j_l=1}^{n^{\mu}} \sum_{k_l=1}^{n^{\nu}}
    \gamma^{((i_l+i_r-1)^2+(j_l-j_r)^2+(k_l-k_r)^2)^{1/2}}
    \right)
    \nonumber
        \end{align}
    \begin{align}
    &\leq \sum_{i_r=1}^{n^{\lambda}/2} \sum_{j_r=1}^{n^{\mu}} \sum_{k_r=1}^{n^{\nu}}
    \nonumber\\ &~~~
    \left( \sum_{v=1}^{\infty} a_1 (3i_r^2+6 i_r v -3i_r+ 3v^2-3v+1)
    \gamma^{i_r+ v-1}
    \right)
    \nonumber\\
    &\leq \sum_{i_r=1}^{n^{\lambda}/2} \sum_{j_r=1}^{n^{\mu}} \sum_{k_r=1}^{n^{\nu}}\left( \sum_{v=1}^{\infty} a_1 (3i_r^2+6 i_r v + 4v^2)\gamma^{i_r+ v-1}
    \right)
    \nonumber\\
    &=\sum_{i_r=1}^{n^{\lambda}/2} \sum_{j_r=1}^{n^{\mu}} \sum_{k_r=1}^{n^{\nu}}\left( a_2 i_r^2\gamma^{i_r}\frac{1}{1-\gamma}
    +a_3 i_r\gamma^{i_r}\frac{1}{(1-\gamma)^2}
    \right.\nonumber\\ &~~~\left.
    + a_4\gamma^{i_r}\frac{(1+\gamma)}{(1-\gamma)^3}
    \right)
    \nonumber\\
    &=\sum_{i_r=1}^{n^{\lambda}/2} \sum_{j_r=1}^{n^{\mu}} \sum_{k_r=1}^{n^{\nu}}\left(a_5 i_r^2\gamma^{i_r}+a_6 i_r\gamma^{i_r} + a_7\gamma^{i_r}
    \right)
    \nonumber\\
    &=\sum_{i_r=1}^{n^{\lambda}/2}n^{\mu+\nu} \left(a_5 i_r^2\gamma^{i_r}+a_6 i_r\gamma^{i_r} + a_7\gamma^{i_r}
    \right)
    \nonumber\\
    &\leq \sum_{i_r=1}^{\infty}n^{\mu+\nu} \left(a_5 i_r^2\gamma^{i_r}+a_6 i_r\gamma^{i_r} + a_7\gamma^{i_r}
    \right)
    \nonumber\\
    &\leq n^{\mu+\nu}\left(a_5\frac{\gamma(1+\gamma)}{(1-\gamma)^3} + a_6\frac{\gamma}{(1-\gamma)^2} + a_7\frac{\gamma}{1-\gamma}
    \right),
    \label{EQ:Upper-Inequality-Exponential}
\end{align}
where $\{a_i\}_{i=1}^7$ are positive constants, independent of
$n$.

Next, using the fact that our 3D random network is transformed to
the regular network with $\log n$ nodes at each left-hand side
vertex and $4\log n$ nodes at each right-hand side vertex,
$T_n^{(2)}$ is finally upper-bounded by
\begin{align}
    T_n^{(2)}
    &\leq 4(\log n)^2 n^{\mu+\nu}
    \nonumber\\ &~~~\cdot
    \left(a_5\frac{\gamma(1+\gamma)}{(1-\gamma)^3} + a_6\frac{\gamma}{(1-\gamma)^2} + a_7\frac{\gamma}{1-\gamma}
    \right), \nonumber
\end{align}
resulting in $T_n^{(2)}=O(n^{\mu+\nu}(\log n)^2)$. Since $T_n\leq
T_n^{(1)}+T_n^{(2)}$ and $T_n^{(1)}=O(n^{\mu+\nu}\log n)$, it
follows that $T_n=O(n^{\mu+\nu}(\log n)^2)$.

If the cuboid is divided into two equal halves by other two cut
planes $L_y$ and $L_z$, then the corresponding upper bounds on the
aggregate throughput are given by $T_n=O(n^{\lambda+\nu}(\log
n)^2)$ and $T_n=O(n^{\lambda+\mu}(\log n)^2)$, respectively, in a
similar fashion. In consequence, by taking the minimum of three
upper bound results, it follows that
$T_n=O(n^{\min\{\mu+\nu,\lambda+\nu,\lambda+\mu\}}(\log
n)^2)=O(n^{\min\{1-\lambda,1-\mu,1-\nu\}}(\log n)^2)$, which
completes the proof of Theorem~\ref{THM:upper_exp}.
\end{IEEEproof}

From Theorems \ref{Theorem:Achievability-Exponential} and
\ref{THM:upper_exp}, the following discussion is made.

\begin{remark}
Under the exponential model in the 3D erasure network, the upper
bound matches the achievable throughput scaling using the
percolation-based 3D highway routing protocol within a
polylogarithmic factor. We also remark that, unlike the case of
Gaussian network models, the use of the hierarchical
cooperation~\cite{OzgurLevequeTse:07} or any sophisticated
multiuser detection scheme is not needed to improve the achievable
throughput scaling.
\end{remark}

\subsection{Upper Bound Under the Polynomial Decay Model}
In this subsection, we present a cut-set upper bound on the
capacity for the polynomial decay model. The proof techniques are
basically similar to those for the exponential decay model.

\begin{theorem} \label{THM:upper_poly}
Suppose a 3D erasure network of size length $n^\lambda \times
n^\mu \times n^\nu$ with unit node density under the polynomial
decay model, where the probability of successful transmission
decays polynomially as in (\ref{Eq:ErasureProb-Polynomial}). Then,
when $\alpha>3$, the total throughput $T_n$ is upper-bounded by
\begin{align}
  T_n=O(n^{\min\{1-\lambda,1-\mu,1-\nu\}}(\log n)^2). \nonumber
\end{align}
\end{theorem}

\begin{IEEEproof}
Let us first assume a regular network having one node at each unit
cube vertex. We start by considering the cut plane $L_x$. Suppose
that the left-hand side nodes are located at positions
$(-i_l+1,j_l,k_l)$ and those on the right-hand side are located at
positions $(i_r,j_r,k_r)$. Substituting
(\ref{Eq:ErasureProb-Polynomial}) into (\ref{Eq:Tn2}), $T_n^{(2)}$
is then bounded by
\begin{align}
    &T_n^{(2)}
    \nonumber\\
    &\leq \sum_{i\in S_L}\sum_{k\in D_L}\frac{1}{d_{ki}^{\alpha}}
    \nonumber\\
    &\leq \sum_{i_l=1,i_r=1}^{n^{\lambda}/2} \sum_{j_l=1,j_r=1}^{n^{\mu}}
    \nonumber\\ &~~~
    \sum_{k_l=1,k_r=1}^{n^{\nu}}
    \frac{1}{((i_l+i_r-1)^2+(j_l-j_r)^2+(k_l-k_r)^2)^{\alpha/2}}
    \nonumber\\
    &\leq \sum_{i_r=1}^{n^{\lambda}/2}\sum_{j_r=1}^{n^{\mu}}\sum_{k_r=1}^{
    n^{\nu}}
    \nonumber\\ &~
    \left(\! \sum_{i_l=1}^{n^{\lambda}/2} \sum_{j_l=1}^{n^{\mu}}\sum_{k_l=1}^{n^{\nu}}
    \frac{1}{((i_l+i_r-1)^2+(j_l-j_r)^2+(k_l-k_r)^2)^{\alpha/2}}\!\!\right)
    \nonumber\\
    &\leq \sum_{i_r=1}^{n^{\lambda}/2}\sum_{j_r=1}^{n^{\mu}}\sum_{k_r=1}^{n^{\nu}}
    \left(
    \sum_{v=1}^{\infty} \frac{a_1(3i_r^2+6i_r v - 3i_r+3 v^2-3v+1)}{(i_r+v-1)^\alpha}
    \right) \nonumber\\
    &\leq \sum_{i_r=1}^{n^{\lambda}/2}\sum_{j_r=1}^{n^{\mu}}\sum_{k_r=1}^{n^{\nu}}
    \left(
    \sum_{v=1}^{\infty} \frac{a_1(3i_r^2+6i_r v+4 v^2)}{(i_r+v-1)^\alpha}
    \right) \nonumber\\
    &\mathop  \leq \limits^{(a)}  \sum_{i_r=1}^{n^{\lambda}/2}\sum_{j_r=1}^{n^{\mu}}\sum_{k_r=1}^{n^{\nu}}
    \left(
    \sum_{v=1}^{\infty} \frac{a_1(3i_r^2+6i_r v+4 v^2)}{i_r^\alpha v^\alpha}
    \right)    \nonumber
    \end{align}
    \begin{align}
    &=
    \sum_{i_r=1}^{n^{\lambda}/2}\sum_{j_r=1}^{n^{\mu}}\sum_{k_r=1}^{n^{\nu}}
    \nonumber\\&~~~
    \left(
    \frac{3a_1}{i_r^{\alpha-2}}\sum_{v=1}^{\infty}\frac{1}{v^{\alpha}}
    + \frac{6a_1}{i_r^{\alpha-1}}\sum_{v=1}^{\infty}\frac{1}{v^{\alpha-1}}
    + \frac{4a_1}{i_r^{\alpha}}\sum_{v=1}^{\infty}\frac{1}{v^{\alpha-2}}
    \right)
    \nonumber\\
    &\leq
    n^{\mu+\nu} \sum_{i_r=1}^{\infty}\left(
    \frac{a_2}{i_r^{\alpha-2}}+\frac{a_3}{i_r^{\alpha-1}}+\frac{a_4}{i_r^{\alpha}}
    \right),
    \label{EQ:Upper-Inequality-Polynomial}
\end{align}
where $\{a_i\}_{i=1}^4$ are positive constants, independent of
$n$. Here, (a) follows from the fact that $i_r v\leq i_r + v-1$.

Next, by using the regular network with $\log n$ nodes at each
left-hand side vertex and $4\log n$ nodes at each right-hand side
vertex, $T_n^{(2)}$ is finally upper-bounded by
\begin{align}
    T_n^{(2)}
    &\leq 4(\log n)^2 n^{\mu+\nu} \sum_{i_r=1}^{\infty}\left(
    \frac{a_2}{i_r^{\alpha-2}}+\frac{a_3}{i_r^{\alpha-1}}+\frac{a_4}{i_r^{\alpha}}
    \right), \nonumber
\end{align}
thus resulting in $T_n^{(2)}=O(n^{\mu+\nu}(\log n)^2)$ when
$\alpha>3$. Since $T_n\leq T_n^{(1)}+T_n^{(2)}$ and
$T_n^{(1)}=O(n^{\mu+\nu}\log n)$, it follows that
$T_n=O(n^{\mu+\nu}(\log n)^2)$ when $\alpha>3$.

If the cuboid is divided into two equal halves by other two cut
planes $L_y$ and $L_z$, then the corresponding upper bounds are
given by $T_n=O(n^{\lambda+\nu}(\log n)^2)$ and
$T_n=O(n^{\lambda+\mu}(\log n)^2)$, respectively, for $\alpha>3$.
Therefore, by taking the minimum of three upper bound results, we
have $T_n=O(n^{\min\{\mu+\nu,\lambda+\nu,\lambda+\mu\}}(\log
n)^2)=O(n^{\min\{1-\lambda,1-\mu,1-\nu\}}(\log n)^2)$ for
$\alpha>3$, which completes the proof of
Theorem~\ref{THM:upper_poly}.
\end{IEEEproof}

By comparing the results in Theorems
\ref{Theorem:Achievability-Exponential}--\ref{THM:upper_poly}, we
have the following observations.

\begin{remark}
It turns out that the upper bounds for both erasure channel models
are of the same order. Moreover, it is shown that the upper bound
for the polynomial decay model also matches the corresponding
achievable throughput scaling within a polylogarithmic factor when
$\alpha>3$; that is, the routing protocol based on the percolation
theory is order-optimal for all operating regimes under the
exponential decay model and for $\alpha>3$ under the polynomial
decay model.
\end{remark}

\section{Extension to the Dense Network Scenario} \label{SEC:Dense}
So far, we have considered extended networks, where the density of
nodes is fixed and the network volume scales as $n$. In this
section, as another network configuration, we consider a dense
erasure network, where $n$ nodes are uniformly and independently
distributed in a cuboid of unit volume, and show its capacity
scaling laws.

First, we would like to address the Gaussian channel setup. In
extended 3D Gaussian networks, the Euclidean distance between
nodes is increased by a factor of $n^{1/3}$, compared to the dense
network case, and hence for the same transmit powers, the received
powers are all decreased by a certain factor. Equivalently, by
re-scaling space, an extended Gaussian network can be regarded as
a dense Gaussian network with the average per-node power
constraint reduced to a certain factor instead of full power while
the received signal-to-interference-and-noise ratios are
maintained as $\Omega(1)$ (refer to~\cite[Section
V]{OzgurLevequeTse:07} for more details).

In the light of the above observation made in Gaussian networks,
in the dense erasure network, the channel models in
(\ref{Eq:ErasureProb-Exponential}) and
(\ref{Eq:ErasureProb-Polynomial}) need to be changed in such a way
that the distance $d_{ki}$ between nodes $i$ and $k$ is scaled up
to $d_{ki}n^{1/3}$, which results in the same erasure events at
the receiver (i.e., the same received signal power) for both
network configurations. More precisely, in the dense erasure
network, we use the following erasure probabilities
$\epsilon_{ki}=1-\gamma^{d_{ki}n^{1/3}}$ and
$\epsilon_{ki}=1-\frac{1}{(d_{ki}n^{1/3})^{\alpha}}$ for the
exponential and polynomial decay models, respectively.

Now, let us show the achievability result in the dense network. It
is obvious to see that the achievable transmission rates $R(d)$
within distance $d$ are the same as Lemmas~\ref{LEM:Rd}
and~\ref{LEM:Rd2} since the number of required time slots in the
$t$-TDMA scheme has still the same order. Therefore, the
achievable throughput scaling laws for the dense network are the
same as those for the extended network shown in
Theorems~\ref{Theorem:Achievability-Exponential}
and~\ref{Theorem:Achievability-Polynomial}. Let us turn to showing
the upper bound results in the dense network. Since the distance
between nodes is re-scaled by $\frac{1}{n^{1/3}}$, we can
similarly use the bounding technique as in
(\ref{EQ:Upper-Inequality-Exponential}) and
(\ref{EQ:Upper-Inequality-Polynomial}). Thus, the upper bounds for
both exponential and polynomial decay models are the same as
Theorems~\ref{THM:upper_exp} and~\ref{THM:upper_poly},
respectively. In consequence, the capacity scaling laws for the
extended erasure network still hold for the dense erasure network.

\section{Concluding Remarks}\label{SEC:Conclusion}
The capacity scaling was completely characterized for a general 3D
erasure random network using two fundamental path-loss attenuation
models, i.e., the polynomial and exponential decay models for the
erasure probability. For the two erasure models, achievable
throughput scaling laws were derived by introducing the 3D
percolation highway system, where packets are delivered through
the highways in $x$, $y$, and $z$ directions. Our result indicated
that the achievable throughput scaling in 3D space is much greater
than that in 2D space since more geographic diversity can be
exploited in 3D space. Cut-set upper bounds were also analyzed
along with the network transformation argument. It turned out that
the upper bounds match the achievable throughput scaling laws
within a polylogarithmic factor for all operating regimes under
the exponential decay model and for $\alpha>3$ under the
polynomial decay model. Further investigation of the capacity
scaling law for 3D erasure networks in the presence of node
mobility remains for future work. Suggestions for further research
also include characterizing the capacity scaling when $\alpha\leq
3$ under the polynomial decay model.

\appendices
\renewcommand\theequation{\Alph{section}.\arabic{equation}}
\setcounter{equation}{0}

\section{Proof of Lemma~\ref{Lemma:MinHighwayPaths}}\label{PF:NumHighwayPaths}
The proof of this lemma essentially follows that of~\cite[Theorem
5]{FranceschettiDouseTseThiran:07} with a slight modification.
Since the event of having a left-to-right crossing of $R_{xz,x}^j$
is an increasing event (see~\cite[Appendix
I]{FranceschettiDouseTseThiran:07} for the explanation of the
increasing event), for all $0<p'<p<1$, we have
\begin{align}
  1-\Pr\{C_{xz,x}^j>\delta\log m_z\}&\leq\left(\frac{p}{p-p'}\right)^{\delta\log
  m_z}
  \nonumber\\
  &~~~\times
  (1-P_{p'}(R_{xz,x}^{j\leftrightarrow})), \nonumber
\end{align}
where $R_{xz,x}^{j\leftrightarrow}$ is the event that there exists
a left-to-right crossing of rectangle $R_{xz,x}^j$ and
$P_p(R_{xz,x}^{j\leftrightarrow})$ is the probability that the
event $R_{xz,x}^{j\leftrightarrow}$ occurs when the probability of
an open edge is given by $p$. For $p>\frac{2}{3}$, we have
\begin{align}
  P_p(R_{xz,x}^{i\leftrightarrow})
  &\geq 1-\frac{4}{3}(m_x+1)e^{-(\kappa\log m_z -\epsilon_m)(-\log(3(1-p))))}
  \nonumber\\
  &\geq
  1-\frac{4}{3}(m_x+1)m_z^{\kappa\log(3(1-p))}(3(1-p))^{-\epsilon_m},
  \nonumber
\end{align}
where the first inequality follows from the same argument as in
the proof of~\cite[Proposition 2]{FranceschettiDouseTseThiran:07}.
By letting $p'=2p-1$, it is seen that $p'>\frac{2}{3}$ due to
$p>\frac{5}{6}$.
From~\cite[Lemma~6]{FranceschettiDouseTseThiran:07}, we thus have
\begin{align}
  &\Pr\{C_{xz,x}^i\leq\delta\log m_z\}
  \nonumber\\
  &\leq\left(\frac{p}{p-p'}\right)^{\delta\log m_z}
  \frac{4}{3}(m_x+1)m_z^{\kappa\log(3(1-p'))}(3(1-p'))^{-\epsilon_m}
  \nonumber\\
  &\leq
  \frac{4}{3}(m_x+1)m_z^{\delta\log\frac{p}{1-p}+\kappa\log(6(1-p))}(6(1-p))^{-\epsilon_m}.
  \nonumber
\end{align}
The probability of having at most $\delta\log m_z$ edge-disjoint
left-to-right crossings in every rectangle $R_{xz,x}^i$ is given
by
\begin{align}\label{Eq:Prob-NumCrossings}
  &\Pr\{N_{xz,x}\leq \delta\log m_z\}
  \nonumber\\
  &= \left(\Pr\{C_{xz,x}^i\leq\delta\log m_z\}\right)^{\frac{m_z}{\kappa\log m_z-\epsilon_m}}
  \nonumber\\
  &\leq\left(
  \frac{4}{3}(m_x+1)m_z^{\delta\log\frac{p}{1-p}+\kappa\log(6(1-p))}
  \right.
  \nonumber\\
  &~~~
  \left.
  \times(6(1-p))^{-\epsilon_m}
  \right)^{\frac{m_z}{\kappa\log m_z-\epsilon_m}}
  \nonumber\\
  &=\left(
  \frac{4}{3}((m_z)^{\lambda/\nu}\frac{1}{c^{\nu/\lambda}}+1)m_z^{\delta\log\frac{p}{1-p}+\kappa\log(6(1-p))}
  \right.
  \nonumber\\
  &~~~
  \left.
  \times(6(1-p))^{-\epsilon_m}
  \right)^{\frac{m_z}{\kappa\log m_z-\epsilon_m}}.
\end{align}
The right-hand side of (\ref{Eq:Prob-NumCrossings}) tends to zero
if
\begin{align}\label{Eq:NumCrossings-Condition}
  \frac{\lambda}{\nu}+\delta\log\frac{p}{1-p}+\kappa\log(6(1-p)) < -1
\end{align}
since $\lim_{x\rightarrow\infty}\left(\frac{1}{x}\right)^x=0$. If
the following inequality is fulfilled, then we can choose
sufficiently small $\delta$ such that
(\ref{Eq:NumCrossings-Condition}) is satisfied:
\begin{align}
  1+\frac{\lambda}{\nu}+\kappa\log(6(1-p)) < 0. \nonumber
\end{align}
In a similar way, if $1+\frac{\nu}{\lambda}+\kappa\log(6(1-p)) <
0$, then one can show that $\Pr\{N_{xz,z}\leq \delta\log
m_x\}\rightarrow 0$ by choosing $\delta$ small enough to satisfy
$\frac{\nu}{\lambda}+\delta\log\frac{p}{1-p}+\kappa\log(6(1-p)) <
-1$. This completes the proof of this lemma.



\begin{thebibliography}{10}
\providecommand{\url}[1]{#1} \csname url@rmstyle\endcsname
\providecommand{\newblock}{\relax}
\providecommand{\bibinfo}[2]{#2}
\providecommand\BIBentrySTDinterwordspacing{\spaceskip=0pt\relax}
\providecommand\BIBentryALTinterwordstretchfactor{4}
\providecommand\BIBentryALTinterwordspacing{\spaceskip=\fontdimen2\font
plus \BIBentryALTinterwordstretchfactor\fontdimen3\font minus
  \fontdimen4\font\relax}
\providecommand\BIBforeignlanguage[2]{{%
\expandafter\ifx\csname l@#1\endcsname\relax
\typeout{** WARNING: IEEEtran.bst: No hyphenation pattern has been}%
\typeout{** loaded for the language `#1'. Using the pattern for}%
\typeout{** the default language instead.}%
\else \language=\csname l@#1\endcsname \fi #2}}

\bibitem{LoLawJacobsson:WC13}
A.~Lo, Y.~W. Law, and M.~Jacobsson, ``A cellular-centric service
architecture
  for machine-to-machine ({M2M}) communications,'' \emph{IEEE Wireless
  Commun.}, vol.~20, no.~5, pp. 143--151, Oct. 2013.

\bibitem{GuptaKumar:00}
P.~Gupta and P.~R. Kumar, ``The capacity of wireless networks,''
\emph{{IEEE}
  Trans. Inform. Theory}, vol.~46, no.~2, pp. 388--404, Mar. 2000.

\bibitem{D.Knuth:76}
D.~E. Knuth, ``Big omicron and big omega and big theta,''
\emph{ACM SIGACT
  News}, vol.~8, no.~2, pp. 18--24, Apr.-Jun. 1976.

\bibitem{FranceschettiDouseTseThiran:07}
M.~Franceschetti, O.~Dousse, D.~N.~C. Tse, and P.~Thiran,
``Closing the gap in
  the capacity of wireless networks via percolation theory,'' \emph{{IEEE}
  Trans. Inform. Theory}, vol.~53, no.~3, pp. 1009--1018, Mar. 2007.

\bibitem{GuptaKumar:03}
P.~Gupta and P.~R. Kumar, ``Towards an information theory of large
networks: an
  achievable rate region,'' \emph{{IEEE} Trans. Inform. Theory}, vol.~49,
  no.~8, pp. 1877--1894, Aug. 2003.


\bibitem{ShinChungLee:TIT13}
W.-Y. Shin, S.-Y. Chung, and Y.~H. Lee, ``Parallel opportunistic
routing in
  wireless networks,'' \emph{{IEEE} Trans. Inform. Theory}, vol.~59, no.~10,
  pp. 6290--6300, Oct. 2013.

\bibitem{NebatCruzBhardwaj:09}
Y.~Nebat, R.~L. Cruz, and S.~Bhardwaj, ``The capacity of wireless
networks in
  nonergodic random fading,'' \emph{{IEEE} Trans. Inform. Theory}, vol.~55,
  no.~6, pp. 2478--2493, June 2009.

\bibitem{ElGamalMammenPrabhakarShah:06}
A.~{El Gamal}, J.~Mammen, B.~Prabhakar, and D.~Shah, ``Optimal
throughput-delay
  scaling in wireless networks--{P}art {I}: {T}he fluid model,'' \emph{{IEEE}
  Trans. Inform. Theory}, vol.~52, no.~6, pp. 2568--2592, June 2006.

\bibitem{NeelyModiano:05}
M.~J. Neely and E.~Modiano, ``Capacity and delay tradeoffs for ad
hoc mobile
  networks,'' \emph{{IEEE} Trans. Inform. Theory}, vol.~51, no.~6, pp.
  1917--1937, June 2005.

\bibitem{OzgurLevequeTse:07}
A.~{\"O}zg{\"u}r, O.~L{\'e}v{\^e}que, and D.~N.~C. Tse,
``Hierarchical
  cooperation achieves optimal capacity scaling in ad hoc networks,''
  \emph{{IEEE} Trans. Inform. Theory}, vol.~53, no.~10, pp. 3549--3572, Oct.
  2007.

\bibitem{GhaderiXieShen:TIT09}
J.~Ghaderi, L.-L. Xie, and X.~Shen, ``Hierarchical cooperation in
ad hoc
  networks: Optimal clustering and achievable throughput,'' \emph{{IEEE} Trans.
  Inform. Theory}, vol.~55, no.~8, pp. 3425--3436, Aug. 2009.

\bibitem{NiesenGuptaShah:09}
U.~Niesen, P.~Gupta, and D.~Shah, ``On capacity scaling in
arbitrary wireless
  networks,'' \emph{{IEEE} Trans. Inform. Theory}, vol.~55, no.~9, pp.
  3959--3982, Sept. 2009.

\bibitem{GrossglauserTse:02}
M.~Grossglauser and D.~N.~C. Tse, ``Mobility increases the
capacity of ad hoc
  wireless networks,'' \emph{{IEEE/ACM} Trans. Networking}, vol.~10, no.~4, pp.
  477--486, Aug. 2002.

\bibitem{CadambeJafar:08}
V.~R. Cadambe and S.~A. Jafar, ``Interference alignment and
degrees of freedom
  of the {$K$}-user interference channel,'' \emph{{IEEE} Trans. Inform.
  Theory}, vol.~54, no.~8, pp. 3425--3441, Aug. 2008.

\bibitem{ZhangXuWangGuizani:TC10}
G.~Zhang, Y.~Xu, X.~Wang, and M.~Guizani, ``Capacity of hybrid
wireless
  networks with directional antenna and delay constraint,'' \emph{{IEEE} Trans.
  Commun.}, vol.~58, no.~7, pp. 2097--2106, July 2010.

\bibitem{LiZhangFang:TMC11}
P.~Li, C.~Zhang, and Y.~Fang, ``The capacity of wireless ad hoc
networks using
  directional antennas,'' \emph{{IEEE} Trans. Mobile Comput.}, vol.~10, no.~10,
  pp. 1374--1387, Oct. 2011.

\bibitem{YoonShinJeon:ISIT14}
J.~Yoon, W.-Y. Shin, and S.-W. Jeon, ``Elastic routing in wireless
networks
  with directional antennas,'' in \emph{Proc. IEEE Int. Symp. Inf. Theory
  (ISIT)}, Honolulu, HI, Jun./Jul. 2014, pp. 1001--1005.

\bibitem{O.Dousse:INFOCOM02}
O.~Dousse, P.~Thiran, and M.~Hasler, ``Connectivity in ad-hoc and
hybrid
  networks,'' in \emph{Proc. {IEEE} {INFOCOM}}, New York, NY, June 2002, pp.
  1079--1088.

\bibitem{KulkarniViswanath:03}
S.~R. Kulkarni and P.~Viswanath, ``Throughput scaling for
heterogeneous
  networks,'' in \emph{Proc. {IEEE} Int. Symp. Inf. Theory (ISIT)}, Yokohama,
  Japan, Jun./Jul. 2003, p. 452.

\bibitem{KozatTassiulas:03}
U.~C. Kozat and L.~Tassiulas, ``Throughput capacity of random ad
hoc networks
  with infrastructure support,'' in \emph{Proc. {ACM} {MobiCom}}, San Diego,
  CA, Sept. 2003, pp. 55--65.

\bibitem{LiuLiuTowsley:03}
B.~Liu, Z.~Liu, and D.~Towsley, ``On the capacity of hybrid
wireless
  networks,'' in \emph{Proc. {IEEE} {INFOCOM}}, San Francisco, CA, Mar./Apr.
  2003, pp. 1543--1552.

\bibitem{ZemlianovVeciana:05}
A.~Zemlianov and G.~de~Veciana, ``Capacity of ad hoc wireless
networks with
  infrastructure support,'' \emph{{IEEE} J. Select. Areas Commun.}, vol.~23,
  no.~3, pp. 657--667, Mar. 2005.

\bibitem{LiuThiranTowsley:07}
B.~Liu, P.~Thiran, and D.~Towsley, ``Capacity of a wireless ad hoc
network with
  infrastructure,'' in \emph{Proc. {ACM} {M}obi{H}oc}, Montr{\'e}al, Canada,
  Sept. 2007, pp. 239--246.

\bibitem{ShinJeonDevroyeVuChungLeeTarokh:08}
W.-Y. Shin, S.-W. Jeon, N.~Devroye, M.~H. Vu, S.-Y. Chung, Y.~H.
Lee, and
  V.~Tarokh, ``Improved capacity scaling in wireless networks with
  infrastructure,'' \emph{{IEEE} Trans. Inform. Theory}, vol.~57, no.~8, pp.
  5088--5102, Aug. 2011.

\bibitem{DoddavenkatappaChanAnanda:ICST11}
M.~Doddavenkatappa, M.~C. Chan, and A.~L. Ananda, ``Indriya: A
low-cost, 3d
  wireless sensor network testbed,'' in \emph{Lecture Notes of the Institute
  for Computer Sciences, Social-Informatics and Telecommunications
  Engineering}.

\bibitem{LinLeuLiWu:Elsevier2013}
M.-S. Lin, J.-S. Leu, K.-H. Li, and J.-L.~C. Wu, ``Zigbee-based
{Internet of
  Things} in {3D} terrains,'' \emph{Computers \& Electrical Engineering},
  vol.~39, no.~6, pp. 1667--1683, Aug. 2013.

\bibitem{GuptaKumar:01}
P.~Gupta and P.~Kumar, ``Internets in the sky: The capacity of
three
  dimensional wireless networks,'' \emph{Commun. Inf. Syst.}, vol.~1, pp.
  33--50, 2001.

\bibitem{LiPanFang:TN12}
P.~Li, M.~Pan, and Y.~Fang, ``Capacity bounds of three-dimensional
wireless ad
  hoc networks,'' \emph{{IEEE/ACM} Trans. Netw.}, vol.~20, no.~4, pp.
  1304--1315, Aug. 2012.

\bibitem{FranceschettiMiglioreMinero:09}
M.~Franceschetti, M.~D. Migliore, and P.~Minero, ``The capacity of
wireless
  networks: information-theoretic and physical limits,'' \emph{{IEEE} Trans.
  Inform. Theory}, vol.~55, no.~8, pp. 3413--3424, Aug. 2009.

\bibitem{HuWangYangZhangXuGao:TC10}
C.~Hu, X.~Wang, Z.~Yang, J.~Zhang, Y.~Xu, and X.~Gao, ``A geometry
study on the
  capacity of wireless networks via percolation,'' \emph{{IEEE} Trans.
  Commun.}, vol.~58, no.~10, pp. 2916--2925, Oct. 2010.

\bibitem{DanaGowaikarHassibi:06}
A.~Dana, R.~Gowaikar, and B.~Hassibi, ``Capacity of wireless
erasure
  networks,'' \emph{{IEEE} Trans. Inform. Theory}, vol.~32, no. 3, pp. 789--804, Mar.
  2006.

\bibitem{LeeUrbankeBlahut:08}
J. W. Lee, R. L. Urbanke, and R. E. Blahut, ``Turbo codes in
binary erasure channel," \emph{{IEEE} Trans. Inf. Theory},
vol.~54, no.~4, pp. 1765--1773, Apr. 2008.

\bibitem{JaberAndrews:11}
R. G. Jabar and J. G. Andrews, ``A lower bound on the capacity of
wireless erasure networks," \emph{{IEEE} Trans. Inf. Theory},
vol.~57, no.~10, pp. 6502--6513, Oct. 2011.

\bibitem{TulinoVerduCaireShamai:07}
A.~Tulino, S.~Verd{\'u}, G.~Caire, and S.~Shamai, ``The {G}aussian
erasure
  channel,'' in \emph{Proc. Int. Symp. Inf. Theory (ISIT)}, Nice, France,
  June/July 2007, pp. 1721--1725.

\bibitem{VerduWeissman:08}
S.~Verd{\'u} and T.~Weissman, ``The information lost in
erasures,''
  \emph{{IEEE} Trans. Inform. Theory}, vol.~54, no.~11, pp. 5030--5058, Nov.
  2008.

\bibitem{SmithGuptaViswanath:07ISIT}
B.~Smith, P.~Gupta, and S.~Vishwanath, ``Routing is order-optimal
in broadcast
  erasure networks with interference,'' in \emph{Proc. {IEEE} Int. Symp. Inf.
  Theory (ISIT)}, Nice, France, June 2007, pp. 141--145.

\bibitem{ShinKim:11}
W.-Y. Shin and A.~Kim, ``Capacity scaling of
infrastructure-supported erasure
  networks,'' \emph{IEEE Commun. Lett.}, vol.~15, no.~5, pp. 485--487, May
  2011.

\bibitem{SmithVishwanath:06}
B.~Smith and S.~Vishwanath, ``Asymptotic transport capacity of
wireless erasure
  networks,'' in \emph{Proc. 44th Allerton Conf. on Commun., Control, and
  Computing}, Monticello, Illinois, 2006, pp. 27--29.

\bibitem{SmithGuptaViswanath:07}
B.~Smith, P.~Gupta, and S.~Vishwanath, ``Routing versus network
coding in
  erasure networks with broadcast and interference constraints,'' in
  \emph{Proc. {IEEE} Military Commun. Conf. (MILCOM)}, Orlando, FL, Oct. 2007,
  pp. 1--5.

\bibitem{JeongShin:CL13}
C.~Jeong and W.-Y. Shin, ``Capacity scaling of hybrid erasure
networks based on
  polynomial power-law,'' \emph{{IEEE} Commun. Lett.}, vol.~17, no.~5, pp.
  1024--1027, May 2013.

\bibitem{B.Smith:08}
B.~M. Smith, \emph{Capacities of Erasure Networks}.\hskip 1em plus
0.5em minus
  0.4em\relax US: ProQuest, 2008.

\bibitem{CoverThomas:91}
T.~M. Cover and J.~A. Thomas, \emph{Elements of Information
Theory}.\hskip 1em
  plus 0.5em minus 0.4em\relax New York: Wiley, 1991.

\end{thebibliography}
\end{document}